\documentclass[a4paper,fleqn,usenatbib]{mnras}

\usepackage{mathptmx}
\usepackage[T1]{fontenc}
\usepackage{ae,aecompl}

\usepackage{graphicx}	% Including figure files
\usepackage{amsmath}	% Advanced maths commands
\usepackage{amssymb}	% Extra maths symbols

\usepackage{natbib}
\usepackage{booktabs}
\usepackage{longtable}
\usepackage[titletoc,title]{appendix}
\usepackage{color}
\usepackage{pdflscape}

\newcommand{\hi}{H~{\small I}}
\newcommand{\hii}{H~{\small II}}
\newcommand{\oi}{[O~{\small I}]}

\newcommand{\oiii}{[O~{\small III}]}

\newcommand{\cii}{[C~{\small II}]}
\newcommand{\ci}{[C~{\small I]}}
\newcommand{\nii}{[N~{\small II]}}
\newcommand{\niii}{[N~{\small III]}}

\usepackage{enumitem}

\title[\cii~ as a molecular gas mass tracer in galaxies]{The \cii~ emission as a molecular gas mass tracer in galaxies at low and high redshift}

\author[A. Zanella et al.]{A. Zanella,$^{1}$\thanks{E-mail:azanella@eso.org} E. Daddi,$^{2}$ G. Magdis,$^{3,4}$ T. Diaz Santos,$^{5}$ D. Cormier,$^{2}$ D. Liu,$^{6}$
\newauthor A. Cibinel,$^{7}$ R. Gobat,$^{8,9}$ M. Dickinson,$^{10}$ M. Sargent,$^{7}$ G. Popping,$^{6}$ S. C. Madden,$^{2}$ 
\newauthor M. Bethermin,$^{11}$ T. M. Hughes,$^{12,13,14,15}$ F. Valentino,$^{3,4}$ W. Rujopakarn,$^{16,17,18}$
\newauthor M. Pannella,$^{19}$ F. Bournaud,$^{2}$ F. Walter,$^{6}$ T. Wang,$^{20}$
D. Elbaz,$^{2}$ R. T. Coogan$^{2,7}$
\\
$^{1}$European Southern Observatory, Karl Schwarzschild Stra\ss e 2, 85748 Garching, Germany\\
$^{2}$CEA, IRFU, DAp, AIM, Universit\'e Paris-Saclay, Universit\'e Paris Diderot, Sorbonne Paris Cit\'e, CNRS, F-91191 Gif-sur-Yvette, France \\
$^{3}$Dawn Cosmic Center, Niels Bohr Institute, University of Copenhagen, Juliane Maries Vej 30, DK-2100 Copenhagen, Denmark \\
$^{4}$Dark Cosmology Centre, Niels Bohr Institute, University of Copenhagen, Juliane Maries Vej 30, DK-2100 Copenhagen, Denmark \\
$^{5}$N\'ucleo de Astronom\'ia de la Facultad de Ingenier\'ia, Universidad Diego Portales, Av. Ej\'ercito Libertador 441, Santiago, Chile \\
$^{6}$Max Planck Institute for Astronomy, Konigstuhl 17, D-69117 Heidelberg, Germany \\
$^{7}$Astronomy Centre, Department of Physics and Astronomy, University of Sussex, Brighton, BN1 9QH, UK \\
$^{8}$School of Physics, Korea Institute for Advanced Study, Hoegiro
85, Dongdaemun-gu, Seoul 02455, Korea \\
$^{9}$Instituto de F\' isica, Pontificia Universidad Cat\' olica de
Valpara\' iso, Casilla 4059, Valpara\' iso, Chile\\
$^{10}$National Optical Astronomy Observatory, Tucson, AZ 85719, USA \\
$^{11}$Aix Marseille Univ, CNRS, LAM, Laboratoire d'Astrophysique de Marseille, Marseille, France \\
$^{12}$Instituto de F\'{i}sica y Astronom\'{i}a, Universidad de
Valpara\'{i}so, Avda. Gran Breta\~{n}a 1111, Valpara\'{i}so, Chile \\
$^{13}$CAS Key Laboratory for Research in Galaxies and Cosmology,
Department of Astronomy, University of Science and Technology of
China,\\ Hefei 230026, China\\
$^{14}$School of Astronomy and Space Science, University of Science and
Technology of China, Hefei 230026, China\\
$^{15}$Chinese Academy of Sciences South America Center for Astronomy,
China-Chile Joint Center for Astronomy, Camino El Observatorio 1515,\\
Las Condes, Santiago, Chile\\
$^{16}$Department of Physics, Faculty of Science, Chulalongkorn University, 254 Phayathai Road, Pathumwan, Bangkok 10330, Thailand \\
$^{17}$National Astronomical Research Institute of Thailand (Public Organization), Don Kaeo, Mae Rim, Chiang Mai 50180, Thailand \\
$^{18}$Kavli Institute for the Physics and Mathematics of the Universe (WPI),The University of Tokyo Institutes for Advanced Study,\\ The
University of Tokyo, Kashiwa, Chiba 277-8583, Japan\\
$^{19}$Faculty of Physics, Ludwig-Maximilians Universit\"at, Scheinerstr. 1, 81679 Munich, Germany\\
$^{20}$Institute of Astronomy, the University of Tokyo, and National Observatory Of Japan, Osawa, Mitaka, Tokyo 181-0015, Japan}

\date{Accepted XXX. Received YYY; in original form ZZZ}

\pubyear{2017}

\hypersetup{draft}
\begin{document}
\label{firstpage}
\pagerange{\pageref{firstpage}--\pageref{lastpage}}
\maketitle

\begin{abstract}
We present ALMA Band 9 observations of the \cii 158$\mu$m  emission for a sample of 10 main-sequence galaxies at redshift $z$$\sim$2, with typical stellar masses ($\log$ M$_\star$/M$_\odot \sim$10.0--10.9) and star
formation rates ($\sim$35--115 M$_\odot$ yr$^{-1}$). 
Given the strong and well understood evolution of the interstellar medium 
from the present to $z=2$, we investigate the behaviour of the \cii\ emission and empirically
identify its primary driver.
We detect \cii\ from six galaxies (four secure, two tentative) and estimate
ensemble averages including non detections.
The \cii -to-infrared luminosity ratio ($L_{\rm \cii}$/$L_{\rm IR}$)
of our sample is similar to that of local main-sequence
galaxies ($\sim2\times10^{-3}$), and $\sim$ 10 times higher than that
of starbursts. The \cii\ emission has an average spatial extent of 4
-- 7~kpc, consistent with the optical size.
Complementing our sample with literature data, we find
that the \cii~ luminosity correlates with galaxies' molecular gas
mass, with a mean absolute deviation of 0.2 dex and without evident
systematics: the \cii -to-H$_2$ conversion factor ($\alpha_{\rm \cii}
\sim 30$ M$_\odot$/L$_\odot$) is largely independent of galaxies'
depletion time, metallicity, and redshift. \cii~ seems therefore a
convenient tracer to estimate galaxies' molecular gas content regardless of their starburst or main-sequence nature, and extending to metal-poor galaxies at low- and high-redshifts.
The dearth of \cii\ emission reported for $z>6$--7 galaxies might suggest either a high star formation efficiency  or  a small fraction of UV light from star formation reprocessed by dust.
\end{abstract}


%%%%%%%%%%%%%%%%%%%%%%%%%%%%%%%%%%%%%%%%%%%%%%%%%%

%%%%%%%%%%%%%%%%% BODY OF PAPER %%%%%%%%%%%%%%%%%%

\section{Introduction}\label{sec:introduction}

A tight correlation between the star formation rates (SFR)
and stellar masses (M$_\star$) in galaxies seems to be in place both in the local Universe
and at high redshift (at least up to redshift $z \sim 7$,
e.g. \citealt{Bouwens2012}, \citealt{Steinhardt2014}, \citealt{Salmon2015}): the so-called
``main-sequence'' (MS; e.g. \citealt{Noeske2007}, \citealt{Elbaz2007},
\citealt{Daddi2007}, \citealt{Stark2009}, followed by many others). The
normalization of this relation increases with redshift. At fixed
stellar mass ($\sim 10^{10}$ M$_\odot$), $z \sim 1$ galaxies have SFRs comparable to local
Luminous Infrared Galaxies (LIRGs); at $z \sim 2$ their SFR is further
enhanced and they form stars at rates comparable to local Ultra Luminous Infrared Galaxies
(ULIRGs). However, the smooth dynamical disk structure of
high-redshift main-sequence sources, together with the tightness of the SFR --
M$_\star$ relation, disfavour the hypothesis that the intense star
formation activity of these galaxies is triggered by major mergers, as
by contrast happens at
$z = 0$ for ULIRGs (e.g., \citealt{Armus1987}, \citealt{Sanders1996}, \citealt{Bushouse2002}). The high SFRs in the distant
Universe seem instead to be sustained by secular processes (e.g. cold
gas inflows) producing more stable star formation histories (e.g.,
\citealt{Noeske2007}, \citealt{Dave2012}).

Main sequence galaxies are
responsible for $\sim 90$\% of the cosmic star formation rate density (e.g. \citealt{Rodighiero2011},
\citealt{Sargent2012}), whereas the remaining $\sim$ 10\% of the cosmic SFR
density is due to sources strongly deviating from the main sequence, showing
enhanced SFRs and extreme infrared luminosities. Similarly
to local ULIRGs, star formation in these starburst (SB) galaxies is
thought to be ignited by major merger episodes (e.g.,
\citealt{Elbaz2011}, \citealt{Nordon2012}, \citealt{Hung2013},
\citealt{Schreiber2015}, \citealt{Puglisi2017}). Throughout this paper
we will consider as starbursts all the sources that fall $> 4$ times above the main sequence \citep{Rodighiero2011}.

To understand the mechanisms triggering star
formation, it is crucial to know the molecular
gas reservoir in galaxies, which forms the main fuel for star formation (e.g. \citealt{Bigiel2008}), at the peak of the
cosmic star formation history ($z \sim 2$). Due to their high
luminosities, the starbursts have been the
main sources studied for a long time, although they only represent a small
fraction of the population of star-forming galaxies. Only recently it
has been possible to gather large samples of $z \sim$ 1 -- 2 main-sequence
sources and investigate their gas content thanks to their CO and dust emission (e.g. \citealt{Genzel2010},
\citealt{Carilli2013}, \citealt{Tacconi2013}, \citealt{Combes2013}, \citealt{Scoville2015}, \citealt{Daddi2015}, \citealt{Walter2016}, \citealt{Dunlop2017}). 
Observing the CO transitions at higher redshift, however, becomes
challenging since the line luminosity dims with cosmological distance, the contrast against the CMB becomes lower
(e.g. \citealt{DaCunha2013}), and it weakens as metallicity decreases
(as expected at high $z$). Some authors describe the latter effect
stating that a large fraction of molecular gas becomes ``CO dark'',
meaning that the CO line no longer traces H$_2$
(e.g. \citealt{Wolfire2010}, \citealt{Shi2016}, \citealt{Madden2016}, \citealt{Amorin2016},
\citealt{Glover2016}) and therefore the CO luminosity per unit gas
mass is much lower on average for these galaxies.
Similarly, the dust content of galaxies
decreases with metallicity and therefore it might not be a suitable
tracer of molecular gas at high redshift. An alternative possibility is to use 
other rest-frame far-infrared (IR) lines instead. Recently \ci~ has been proposed as
molecular gas tracer (e.g., \citealt{Papadopoulos2004}, \citealt{Walter2011}, \citealt{Bothwell2016},
\citealt{Popping2017}), although it is fainter than many CO transitions and this is still an open
field of research. Alternatively the \cii~ $^2P_{3/2}$ -- $^2P_{1/2}$
transition at 158 $\mu$m might be a promising tool to investigate the gas
physical conditions in the distant Universe (e.g. \citealt{Carilli2013}). 

\cii~ has been identified as one of the brightest fine structure lines emitted from
star-forming galaxies. It has a lower ionization potential than \hi~
(11.3 eV instead of 13.6 eV) and therefore it can be produced in
cold atomic interstellar medium (ISM), molecular, and ionized gas. However, several studies have
argued that the bulk of galaxies' \cii~ emission originates in the external layers of molecular clouds heated by the
far-UV radiation emitted from hot stars with $\gtrsim$ 60 -- 95\% of the total \cii~ luminosity
arising from photodissociation regions (PDRs, e.g. \citealt{Stacey1991},
\citealt{Sargsyan2012}, \citealt{Rigopoulou2014}, \citealt{Cormier2015}, \citealt{Diaz-Santos2017}, \citealt{Croxall2017}). In particular, \cite{Pineda2013} and \cite{Velusamy2014} showed that $\sim$ 75\% of the \cii~ emission in the Milky Way is coming from the molecular gas; this is in good agreement with simulations showing that 60\% -- 85\% of the \cii~ luminosity emerges from the molecular phase (\citealt{Olsen2017}, \citealt{Accurso2017b}, \citealt{Vallini2015}). There are also observational and theoretical models suggesting that \cii~ is a good tracer of the putative ``CO dark'' gas. 
The main reason for this is the fact that in the outer regions of
molecular clouds, where the bulk of the gas-phase carbon resides,
H$_2$ is shielded either by dust or self-shielded from UV
photodissociation, whereas CO is more easily photodissociated into C
and C$^+$. This H$_2$ is therefore not traced by CO, but it mainly
emits in \cii~  (e.g. \citealt{Maloney1988}, \citealt{Stacey1991}, \citealt{Madden1993}, \citealt{Poglitsch1995}, \citealt{Wolfire2010}, \citealt{Pineda2013}, \citealt{Nordon2016}, \citealt{Fahrion2017}, \citealt{Glover2016}). Another advantage of using the \cii~ emission line is the fact that it possibly traces also molecular gas with moderate density. In fact, the critical density needed to excite the \cii~ emitting level through electron impacts is $>$ 10 particle/cc ($\sim$ 5 - 50 cm$^{-3}$). For comparison, the critical density needed for CO excitation is higher ($\sim$ 1000 H/cc), so low-density molecular gas can emit \cii, but not CO (e.g. \citealt{Goldsmith2012}, \citealt{Narayanan2017}). This could be an important contribution given the fact that $\sim$ 30\% of the molecular gas in high-redshift galaxies has a density $<$ 50 H/cc \citep{Bournaud2017}, although detailed simulations of the \cii~ emission in turbulent disks are still missing and observational constraints are currently lacking.

The link between the \cii~
emission and star-forming regions is further highlighted by the well
known relation between the \cii~ and IR luminosities
($L_{\mathrm{\cii}}$ and $L_{\mathrm{IR}}$ respectively,
e.g. \citealt{deLooze2010}, \citealt{deLooze2014},
\citealt{Popping2014}, \citealt{Herrera-Camus2015},
\citealt{Popping2016}, \citealt{Olsen2016}, \citealt{Vallini2016}), since the IR
luminosity is considered a good indicator of the SFR
\citep{Kennicutt1998}. However, this relation is not unique and different galaxies show distinct $L_{\mathrm{\cii}}/L_{\mathrm{IR}}$ ratios. In fact,
\begin{landscape}
\begin{figure}
\includegraphics[width=1.35\textwidth]{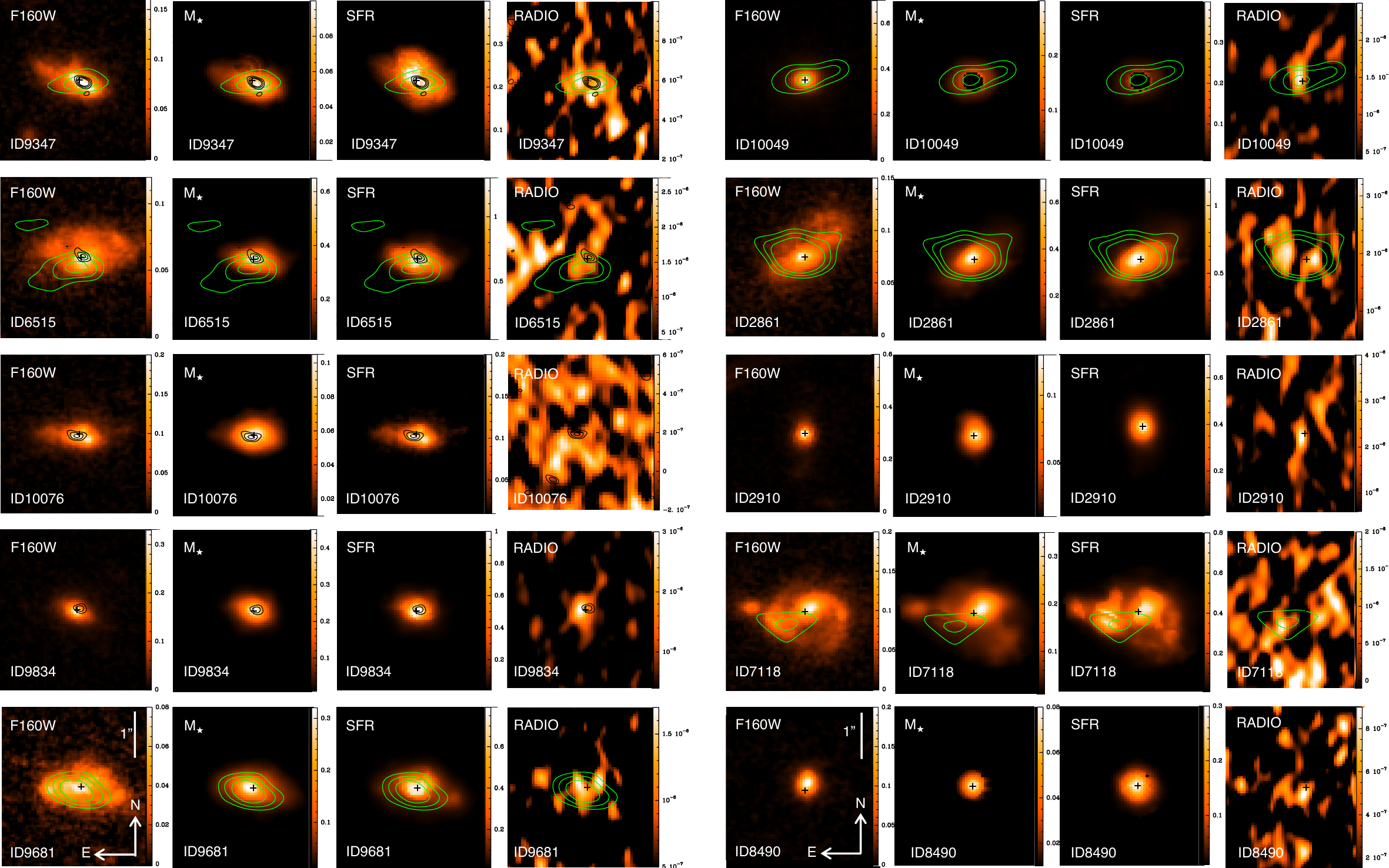}
\caption{\textit{HST} and ALMA observations of our sample
  galaxies. For each source we show the \textit{HST}/WFC3 image
  taken with the F160W filter, the stellar mass map, the star
  formation rate map, and the radio observations taken with VLA. The
  overplotted black contours, when present, show the $> 3\sigma$ \cii~
  emission. The green contours indicate the $> 3\sigma$ 850 $\mu$m
  continuum. The color scale in all panels is linear and it is chosen
  to show galaxies' features at best. The units of the color bars are
  the following: counts s$^{-1}$ for F160W, 10$^{9}$ M$_\odot$ for the
  stellar mass maps, M$_\odot$ yr$^{-1}$ for the SFR maps, and Jy for
  the radio.}
\label{fig:hst_panels}
\end{figure}
\end{landscape} 
\noindent in the local Universe main-sequence sources show a constant
$\langle L_{\mathrm{\cii}}/L_{\mathrm{IR}}\rangle \sim $ 0.002 -- 0.004, although with
substantial scatter (e.g., \citealt{Stacey1991},
\citealt{Malhotra2001}, \citealt{Stacey2010}; \citealt{Cormier2015},
\citealt{Smith2017}, \citealt{Diaz-Santos2017}). Whereas when
including also local starburst galaxies (LIRGs and ULIRGs) with
$L_{\mathrm{IR}} > 10^{11}$ L$_\odot$, the \cii$/$IR luminosity ratio
drops significantly by up to an order of magnitude (e.g. \citealt{Malhotra1997}, \citealt{Stacey2010},
\citealt{Diaz-Santos2013}, \citealt{Farrah2013}, \citealt{Magdis2014}). These sources are
usually referred to as ``\cii~ deficient'' with respect to main-sequence
galaxies. It has been shown that not only
the \cii~ emission drops, but also other
far-IR lines tracing both PDRs and \hii~
regions (e.g. \oi 145 $\mu$m, \nii 122 $\mu$m, \oiii 88 $\mu$m, \oi
63 $\mu$m, \niii 57 $\mu$m, \citealt{Gracia-Carpio2011}, \citealt{Zhao2013}, \citealt{Diaz-Santos2017}) show a deficit when starbursts are considered. This is likely related to the enhanced star formation
efficiency (SFE = SFR/$M_{\mathrm{mol}}$) of starbursts with respect to local main-sequence
galaxies, consistent with the results by \cite{Daddi2010} and
\cite{Genzel2010}. This relation between the $L_{\mathrm{\cii}}/L_{\mathrm{IR}}$ and galaxies' SFE could be due to the fact that the average properties of the interstellar medium in main-sequence and starburst sources are significantly different: the highly compressed and more efficient star formation in starburst could enhance the ionization parameters and drive to lower line to continuum ratios \citep{Gracia-Carpio2011}.
At high redshift, observations become more
challenging, mainly due to the fainter fluxes of the targets: so far $z > 1$ studies
have mainly targeted IR selected sources (e.g., the most luminous
sub-millimeter galaxies and quasars), whereas measurements for IR
fainter main-sequence targets are still limited (e.g., \citealt{Stacey2010},
\citealt{Hailey-Dunsheath2010}, \citealt{Ivison2010}, \citealt{Swinbank2012}, \citealt{Riechers2014},
\citealt{Magdis2014}, \citealt{Huynh2014}, \citealt{Brisbin2015}). Therefore it is not clear yet if high-$z$ main-sequence
galaxies, which have similar SFRs as (U)LIRGs, are expected to be \cii~ deficient.  With our sample we start to push the limit of current observations up to redshift $z \sim 2$.

The goal of this paper is to understand whether main-sequence, $z \sim
2$ galaxies are \cii~ deficient and investigate what are the main
physical parameters the \cii~ emission line is sensitive to. Interestingly we find that
its luminosity traces galaxies' molecular gas mass and could therefore be used as an alternative to other
proxies (e.g. CO, [CI], or dust emission). Given its brightness and the fact that it remains luminous at low metallicities where the CO largely fades, this
emission line might become a valuable resource to explore the galaxies'
gas content at very high redshift. Hence understanding the \cii~
behaviour in $z \sim 2$ main-sequence galaxies, whose physical properties are
nowadays relatively well constrained, will lay the ground for future
explorations of the ISM at higher redshift.

The paper is structured as follows: in Section \ref{sec:data}
we present our observations, sample selection, and data analysis; in
Section \ref{sec:results} we discuss our results; in Section
\ref{sec:summary} we conclude and summarize. Throughout the paper we
use a flat $\Lambda$CDM cosmology with $\Omega_{\mathrm{m}} = 0.3$,
$\Omega_{\mathrm{\Lambda}} = 0.7$, and $H_{0} = 70\, \mathrm{km\, s^{-1}
Mpc^{-1}}$. We assumed a \cite{Chabrier2003} initial mass function
(IMF) and, when necessary, we accordingly converted literature results obtained with
different IMFs.

\section{Observations and data analysis}
\label{sec:data}
In this Section we discuss how we selected the sample and we present
our ALMA observations together with available ancillary data. We also
report the procedure we used to estimate the \cii~ and continuum flux
of our sources. Finally, we describe the literature data that we used to complement our observations, for which full details are given in Appendix.

\subsection{Sample selection and ancillary data}
\label{subsec:sample}

To study the ISM properties of high-redshift main-sequence galaxies, we selected targets in the GOODS-S field (\citealt{Giavalisco2004}, \citealt{Nonino2009}), which benefits from extensive multi-wavelength coverage.

Our sample galaxies were selected on the basis of the following
criteria: 1) having spectroscopic redshift in the range
$1.73 < z < 1.94$ to target the \cii~ emission line in ALMA Band 9. We
made sure that the selected galaxies would have been observed in a
frequency region of Band 9 with good atmospheric transmission. Also, to minimize  overheads, we selected
our sample so that multiple targets could be observed with the same
ALMA frequency setup; 2) being detected in the
available \textit{Herschel} data; 
3) having SFRs
and M$_\star$ typical of main-sequence galaxies at this redshift, as defined by
\citet[they all have  sSFR/sSFR$_\mathrm{MS} <$ 1.7]{Rodighiero2014};
 4) having undisturbed morphologies, with no clear indications of ongoing mergers, as
inferred from the visual inspection of \textit{HST} images. Although
some of the optical images of these galaxies might look disturbed, their
stellar mass maps are in general smooth (Figure \ref{fig:hst_panels}), indicating that the
irregularities visible in the imaging are likely due to star-forming
clumps rather than major mergers (see, e.g., \citealt{Cibinel2015}).

Our sample therefore consists of 10 typical star-forming, main-sequence
galaxies at redshift $1.73 \leq z \leq 1.94$. Given the high
ionization lines present in its optical
spectrum, one of them (ID10049) appears to host an active galactic
nucleus (AGN). This source was not detected in \cii~ and retaining it or not in our final sample does not impact the implications of this work.

Deep Hubble Space Telescope (\textit{HST})
observations at optical
(\textit{HST}/ACS F435W, F606W, F775W, F814W, and F850LP filters) and near-IR (\textit{HST}/WFC3 F105W, F125W, and F160W filters) wavelengths are available from the
CANDELS survey (\citealt{Koekemoer2011}, \citealt{Grogin2011}). \textit{Spitzer} and \textit{Herschel} mid-IR
and far-IR photometry in the wavelength range 24 $\mu$m -- 500 $\mu$m is
also available (\citealt{Elbaz2011}, \citealt{Wang2017}). Finally, radio observations at $\sim$ 5 cm (6
GHz) were taken with the Karl G. Jansky Very Large Array (VLA) with 0.3'' $\times$ 0.6'' resolution \citep{Rujopakarn2016}.

Thanks to these multiwavelength data, we created resolved stellar mass
and SFR maps for our targets, following the method described by
\cite{Cibinel2015}. In brief, we performed pixel-by-pixel spectral
energy distribution (SED) fitting considering all the available \textit{HST}
filters mentioned above, after having convolved all the images with the PSF of the
matched $H_{\mathrm{F160W}}$ band, useful also to increase the signal-to-noise
(S/N). We considered \cite{Bruzual2003}
\begin{landscape}
\begin{figure}
\centering
\includegraphics[width=1.35\textwidth]{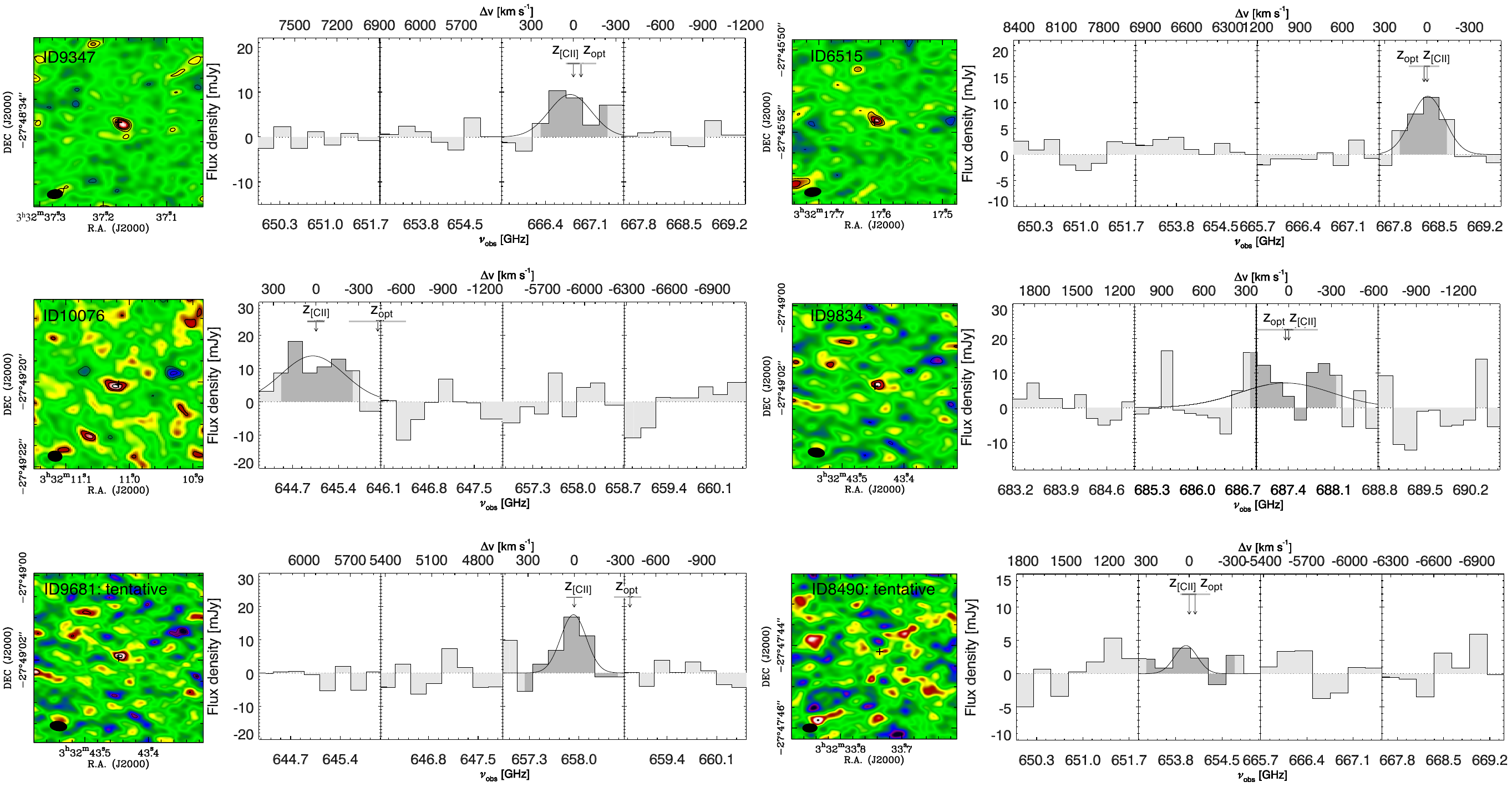}
\caption{ALMA spectra of the \cii~ detections of our
  sample. Left panels: ALMA 2D maps of the \cii~ emission line. The
  black solid and dashed contours indicate respectively the positive and negative 3$\sigma$, 4$\sigma$, and 5$\sigma$ levels. The beam is reported as the black filled
  ellipse. Each stamp has a size of 4'' $\times$ 4''. The black cross
  indicates the galaxy center, as estimated from the \textit{HST} F160W imaging. Some tapering
  has been done for illustrative purposes, although we used the
  untapered maps for the analysis. Right panels: 1D spectra of the
  \cii~ detected sources extracted using a PSF to maximize the S/N
  (notice that in this figure we did not scale the fluxes of the
  spectra extracted with PSF to match those
  obtained when using an exponential function
  with larger size as reported in Table \ref{tab:properties}). The dark grey shaded areas indicate the
  1-$\sigma$ velocity range over which the flux has been measured.
  The frequencies corresponding to the optical and \cii~ redshifts are marked with arrows. The horizontal bars indicate the 1$\sigma$
  uncertainty associated to the optical (light gray) and \cii~ (dark
  gray) redshift estimate. For illustrative purposes we also report
  the Gaussian fit of the emission lines: it was not used to estimate
  the line fluxes, but only as an alternative estimate of the galaxies'
  redshift (Section \ref{subsec:cii_measurements}).}
\label{fig:cii_images}
\end{figure}
\end{landscape}
\noindent templates with constant SFR to limit the degeneracy with dust
extinction. We corrected the fluxes for dust extinction 
 following the prescriptions by
\cite{Calzetti2000}. The stellar population age in the models varied
between 100 Myr and 2 Gyr, assuming fixed solar metallicity. In Figure \ref{fig:hst_panels} we show the resulting SFR and stellar mass maps, together
with the \textit{HST} $H_{\mathrm{F160W}}$-band imaging.
The stellar mass
computed summing up all the pixels of our maps is in good agreement with that estimated by \cite{Santini2014} fitting the global ultraviolet (UV) to IR SED (they differ $<$ 30\% with no systematic trends). In the following we use the stellar masses obtained from the global galaxies' SED, but our conclusions would not change considering the estimate from the stellar mass maps instead.

Spectroscopic redshifts for our sources  are all publicly available and were determined in different
ways: 5 of them are from the GMASS survey \citep{Kurk2013}, one from the K20
survey (\citealt{Cimatti2002}, \citealt{Mignoli2005}), 2 were determined by
\cite{Popesso2009} from VLT/VIMOS spectra, one was estimated from our
rest-frame UV Keck/LRIS spectroscopy as detailed below, and one 
had a spectroscopic redshift estimate determined by \cite{Pope2008}
from PAH features in the \textit{Spitzer}/IRS spectrum. 
With the exception of three sources\footnote{ID2910 that had an IRS spectrum, ID10049 that is an AGN, and ID7118 that has a spectrum from the K20 survey and whose redshift was measured from the H$\alpha$ emission line}, all the redshifts were estimated from rest-frame UV absorption lines. This is a notoriously difficult endeavour especially when, given the faint UV magnitudes of the sources, the  signal-to-noise ratio (S/N) of the UV continuum is moderate, as for our targets. 
We note that having accurate spectroscopic redshifts is crucial for data like that presented here: ALMA observations are carried out using four, sometimes adjacent, sidebands (SBs) covering 1.875~GHz each, corresponding to only 800~km~s$^{-1}$ rest-frame in Band~9 (or equivalently $\Delta z=0.008$). This implies that the \cii~ emission line might be outside the covered frequency range for targets with inaccurate spectroscopic redshift. In general we used at least two adjacent SBs (and up to all 4 in one favourable case) targeting, when possible, galaxies at comparable redshifts (Table \ref{tab:log}). 

Given the required accuracy in the redshift estimate, before the finalization of the observational setups, we carefully re-analyzed all the spectra of our targets to check and possibly refine the redshifts already reported in the literature. To this purpose, we applied to our VLT/FORS2 and Keck UV rest-frame spectra the same approach described in \citet[although both the templates we used and the wavelength range of our data are different]{Gobat2017}. Briefly, we modelled the $\sim$ 4000 -- 7000 \AA~ range of the spectra using standard
Lyman break galaxy templates from \cite{Shapley2003}, convolved with a
Gaussian to match the resolution of our observations. The redshifts
were often revised with respect to those published\footnote{At this
  stage we discovered that one of the literature redshifts was
  actually wrong, making \cii\ unobservable in Band~9. This target was
  dropped from the observational setups, and so we ended up observing
  a sample of 10 galaxies instead of the 11 initially allocated to our
  project.} with variations up to $\sim$ a few $\times 10^{-3}$. Our new values, reported in Table \ref{tab:properties}, match those measured in the independent work of \cite{Tang2014} and have formal uncertainties $\lesssim 1$--2$\times10^{-3}$ ($\lesssim$ 100$-200~$km s$^{-1}$), corresponding to an accuracy in the estimate of the \cii~ observed frequency of $\sim$ 0.25 GHz. 

\begin{figure*}
\includegraphics[width=\textwidth]{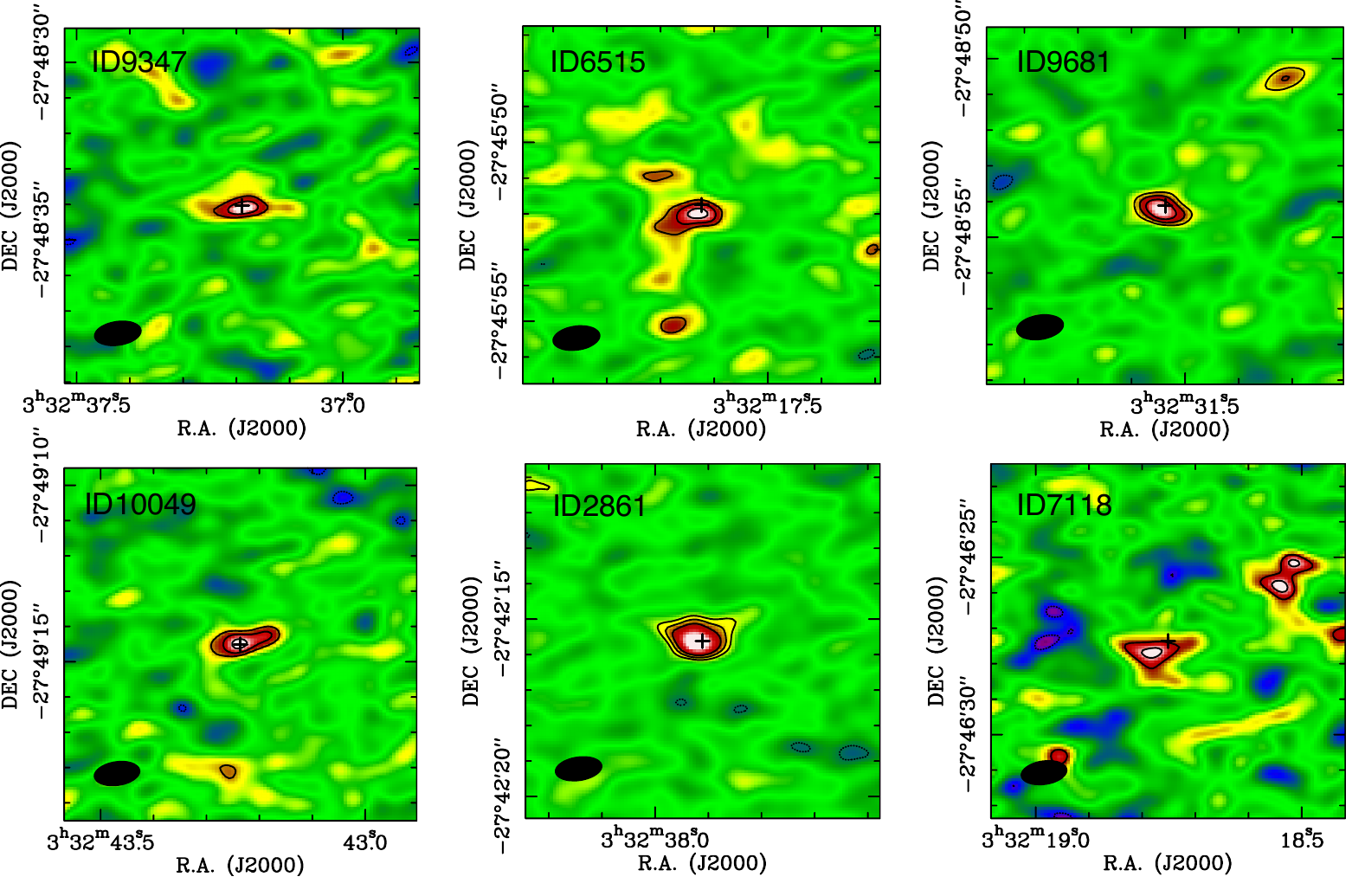}
\caption{ALMA maps of the continuum detections at 850 $\mu$m. The black contours indicate the 3$\sigma$, 4$\sigma$, and 5$\sigma$ levels. The beam is reported as the black filled ellipse. Each stamp has a size of 10'' $\times$ 10''.  The black cross
  indicated the galaxy center, as estimated from the \textit{HST} imaging.  Some tapering
  has been done for illustrative purposes, although we used the
  untapered maps for the analysis.}
\label{fig:850_images}
\end{figure*}

\begin{figure*}
\centering
\includegraphics[width=\textwidth]{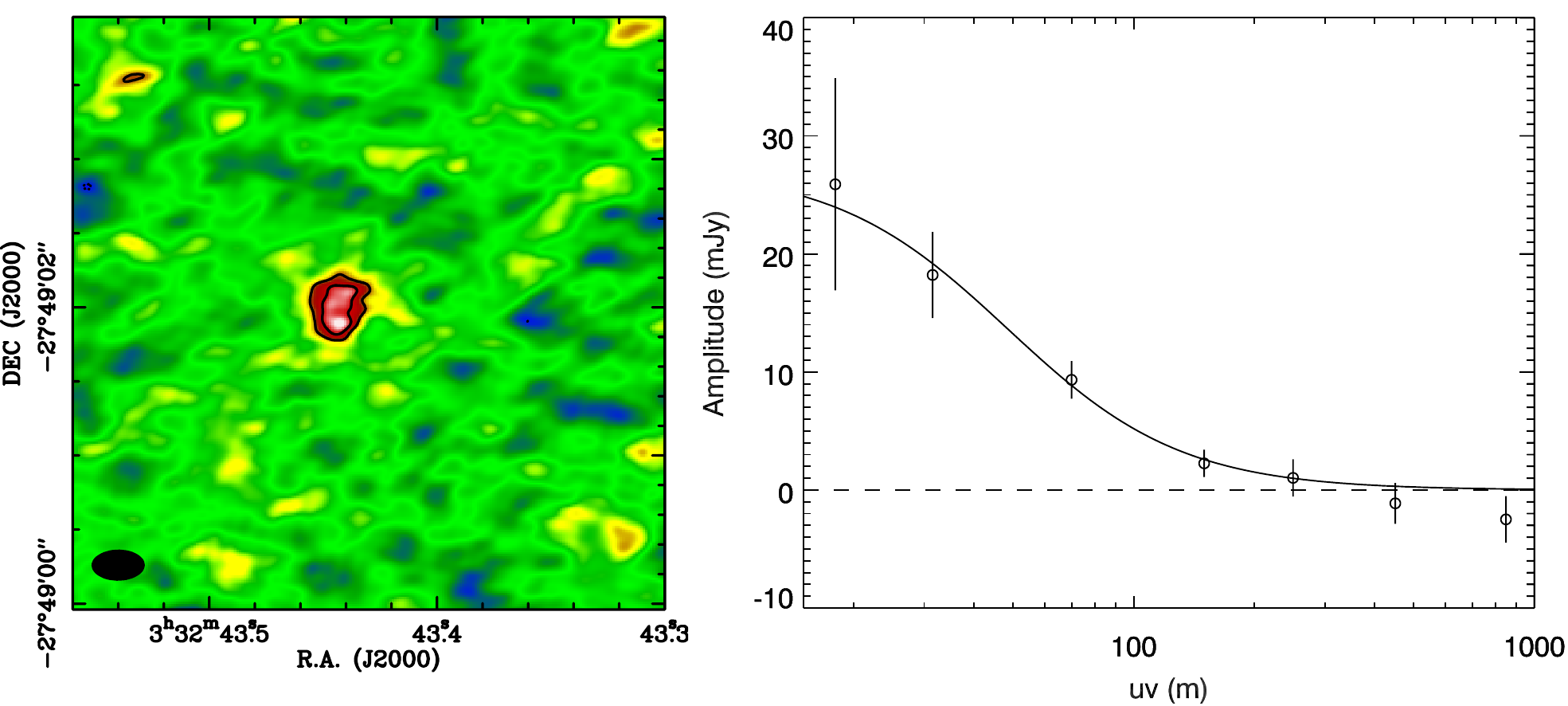}
\caption{Stacking of the four secure \cii~ detections of our
  sample. Left panel: image obtained aligning the four galaxies at
  their \textit{HST} peak positions and stacking their
  visibilities. From 3$\sigma$ and 4$\sigma$ contours are shown. Right
  panel: signal amplitude as a function of the \textit{uv} distance
  (namely the baseline length). We fitted the data with an exponential
  model (black curve). A similar fit is obtained when fitting the data
with a gaussian model with FWHM $\sim$ 0.6''.}
\label{fig:ampl}
\end{figure*}

\begin{figure*}
\includegraphics[width=\textwidth]{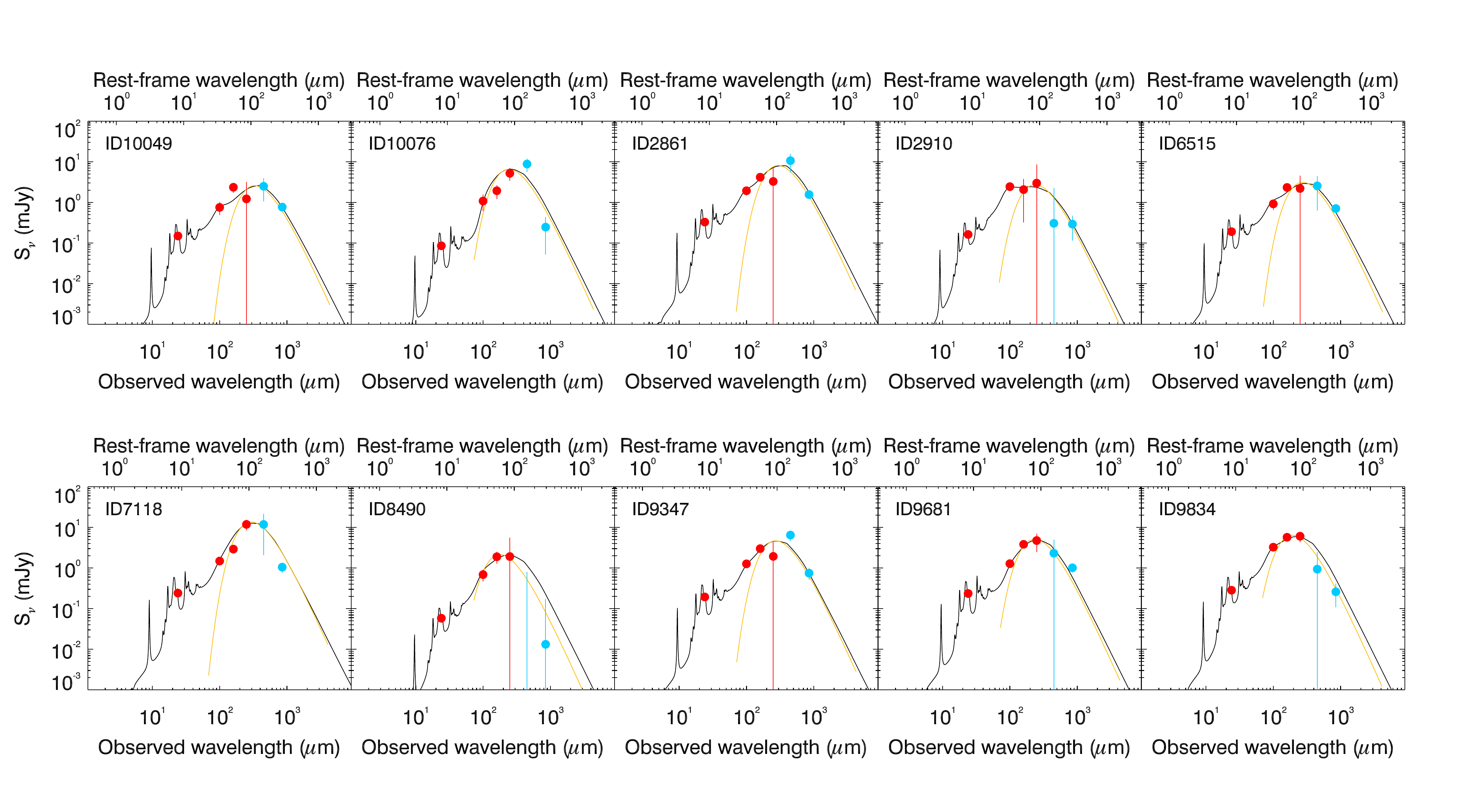}
\caption{Spectral energy distribution fits for our sample
  galaxies. \textit{Herschel} and \textit{Spitzer} measurements are reported as red
  filled circles and the ALMA ones as cyan filled circles. The black curve is the best model fit and the yellow line indicates the best modified black body fit.}
\label{fig:sed}
\end{figure*}

\subsection{Details of ALMA observations}
\label{subsec:details_obs}

We carried out ALMA Band 9 observations for our sample during Cycle 1 (PI: E. Daddi, Project ID: 2012.1.00775.S) with the goal of detecting the \cii~ emission line at rest-frame 158 $\mu$m
($\nu_{\mathrm{rest-frame}} = 1900.54$ GHz) and
the underlying continuum, redshifted in the
frequency range $\nu_{\mathrm{obs}} = 645$ -- $696$ GHz. Currently
this is the largest sample of galaxies observed with ALMA at this redshift with available
\cii~ measurements given
the difficulty to carry out such observations in Band 9. We observed each galaxy, depending on its IR luminosity,
for 8 -- 13 minutes including overheads to reach a homogeneous sensitivity of 1.5 -- 2 mJy/beam over a bandwidth of 350 km s$^{-1}$. 
We set a spectral
resolution of 0.976 MHz (0.45~km~s$^{-1}$ -- later binning the data to substantially lower velocity resolutions) and we requested observations with a spatial
resolution of about 1'' (configuration C32-1) to get integrated
flux measurements of our sources. However, the observations were taken in the C32-3 configuration
with a synthesized beam FWHM $=$ 0.3'' $\times$ 0.2'' and a maximum recoverable scale of
$\sim$ 3.5''. Our sources were therefore resolved. To check if we could still correctly estimate total \cii~
fluxes, we simulated with CASA \citep{McMullin2007} observations in the
C32-3 configuration of extended sources with sizes comparable to those of our galaxies, as detailed in Appendix \ref{app:resolution}. We concluded that, when fitting the sources in the \textit{uv} plane, we
could measure their correct total fluxes, but with substantial losses in terms of effective depth of the data. Figure \ref{fig:sims} in Appendix \ref{app:resolution} shows how the total flux error of a source increases, with respect to the case of unresolved observations,  as a function of its size expressed in units of the PSF FWHM (see also Equation \ref{eq:noise} that quantifies the trend).
Given that our targets are 3 -- 4 times larger than the PSF, we obtained a flux measurement error  5 -- 10 times higher than expected, hence correspondingly lower S/N ratios. The depth of our data, taken with 0.2'' resolution, is therefore equivalent to only 10 -- 30s of integration if taken with 1$''$ resolution. However, when preparing the observations we considered conservative estimates of the \cii~ flux and therefore several targets were detected despite the higher effective noise.

As part of the same ALMA program, besides the Band 9 data, we also requested
additional observations in Band 7 to detect the 850 $\mu$m continuum, which is
important to estimate dust masses for our targets (see Section \ref{subsec:continuum_measurements}). For each galaxy we reached a sensitivity of 140 $\mu$Jy/beam on the continuum, with an integration time of $\sim 2$ minutes on source. The
synthesised beam has FWHM $=$ 1'' $\times$ 0.5'' and the maximum recoverable scale
is $\sim$ 6''.

We note that there is an astrometry offset between our ALMA
observations and the \textit{HST} data released in the GOODS-S
field (Appendix \ref{app:astrometry}). Although it is negligible in right ascension ($\Delta RA =
0.06$''), it is instead significant in declination ($\Delta DEC = -0.2$'',
$> 3\sigma$ significant), in agreement with estimates reported by other studies
(e.g. \citealt{Dunlop2017}, \citealt{Rujopakarn2016},
\citealt{Barro2016}, \citealt{Aravena2016b}, \citealt{Cibinel2017}). We accounted for this offset when interpreting our data by shifting the \textit{HST}
coordinate system to match that of ALMA. In Figure
\ref{fig:hst_panels} we show the astrometry-corrected \textit{HST}
stamps. However, in Table \ref{tab:properties} we report the
uncorrected \textit{HST} coordinates to allow an easier comparison with previous studies. The ALMA target positions are consistent with those from VLA.

\subsection{\cii~ emission line measurements}
\label{subsec:cii_measurements}

The data were reduced with the standard ALMA pipeline based on the CASA software
\citep{McMullin2007}. The calibrated data cubes were then converted to {\em uvfits} format and analyzed with the software GILDAS \citep{Guilloteau2000}.

To create the velocity-integrated \cii~ line maps for our sample
galaxies it was necessary to determine the spectral range over which
to integrate the spectra. This in turn requires a 1D spectrum, that
needs to be extracted at some spatial position and with a source surface brightness distribution model (PSF or extended).  We carried out the following iterative procedure, similar to what described in Daddi et al. (2015 and in preparation) and 
\cite{Coogan2017}.

We fitted, in the \textit{uv} plane, a given source model (PSF, but also Gaussian and exponential profiles, tailored to the \textit{HST} size of the galaxies) to all four sidebands and channel per channel, with fixed spatial position 
determined from the astrometry-corrected \textit{HST} images. We
looked for positive emission line signal in the resulting
spectra. When a signal was present, we averaged the data over the
channels maximizing the detection S/N and we fitted the resulting single channel dataset to obtain the best fitting line spatial position. If this was different from the spatial position of the initial  extraction we proceeded to a new spectral extraction at the new position, and iterate the procedure until convergence was reached. 

\subsubsection{Individual \cii~ detections}
\label{subsec:cii_detections}

Four galaxies converged to secure detections (Figure ~\ref{fig:cii_images}): 
they have emission line significance $> 5 \sigma$ in the optimal channel range. The detections are robust against the model used for the extraction of the 1D spectra: the frequency range used for the lines' identification would not change if we extracted the 1D spectra with a Gaussian or exponential model instead of a PSF. The optimizing spatial positions for spectral extractions were consistent with the \textit{HST} peak positions, typically within the PSF FWHM (Figure ~\ref{fig:cii_images}), and the spectra extracted with Gaussian or exponential models were in any case invariant with respect to such small spatial adjustments.

We estimated the redshift of the four detections in two ways, both giving consistent results (redshift differences $< 0.001$) and similar formal redshift uncertainties: 1) we computed the  signal-weighted average frequency within the line channels, and 2) we fitted the 1D spectrum with a Gaussian function. 
Following \cite{Coogan2017} simulations of a similar line detection procedure, and given the S/N of these detections, we concluded that redshift uncertainties estimated in this way are reliable. 
We compared our redshift estimates for these sources with
those provided by our VLT and Keck data analysis, and in the literature (Section \ref{sec:data}). They generally agree,
with no significant  systematic difference and a median absolute deviation (MAD) of 200~km~s$^{-1}$ (MAD$_{\rm z}=0.002$). 
This accuracy is fully within the expected uncertainties of both our optical and \cii\ redshift (see Table \ref{tab:properties}), thus increasing the reliability of the detections considering that the line search was carried out over a total $\Delta z=0.035$. 

Given the fact that our sources are extended, we estimated their total \cii~ flux by fitting their average
emission line maps in the \textit{uv} plane with exponential models (whereas by using a PSF model instead we would have underestimated the fluxes). We used the following procedure.
Our sample is composed of disk-like galaxies as shown in Figure \ref{fig:hst_panels}. Although in some cases (e.g. ID7118) some clumps of star formation are visible both in the \textit{HST} imaging and in the spatially resolved SFR maps, the resolved stellar mass maps are smooth, as expected for unperturbed sources, and mainly show the diffuse disk seen also in our ALMA observations. We therefore determined the size of the galaxy disks by fitting the stellar mass maps with an exponential profile \citep{Freeman1970}, using the GALFIT algorithm \citep{Peng2010}. We checked that there were not structured residuals when subtracting the best-fit model from the stellar mass maps. We then extracted the \cii~ flux by
fitting the ALMA data in the $uv$ plane, using the Fourier Transform of the 2D
exponential model, with the GILDAS task \texttt{uv\_fit}. We fixed the size
and center of the model on the basis of the effective radius and peak coordinates derived from
the optical images, corrected for the astrometric offset determined as
in Appendix \ref{app:astrometry}. As a result, we obtained the total
\cii~ flux of our sources. Given the 
larger uncertainties associated to extended source models with respect to the PSF case (Appendix \ref{app:astrometry}), this procedure returns  $> 3\sigma$ total flux measurements for the four sources (even if original detections were $> 5\sigma$). We checked that fluxes and
uncertainties determined with the \texttt{uvmodelfit} task provided by CASA
would give consistent results. We also checked the robustness
of our flux measurements against the assumed functional form of the
model: fitting the data with a Gaussian profile instead of an
exponential would give consistent \cii~
fluxes. Finally, we verified that the uncertainties associated to the flux measurement in each channel are consistent with the channel to channel fluctuations, after accounting for the continuum emission and excluding emission lines.

\begin{table*}
\center
\caption{Log of the observations}
\label{tab:log}
\begin{tabular}{c c c c c c c c}
\toprule
\midrule
ID & Date & $z_{\mathrm{SB1}}$ & $z_{\mathrm{SB2}}$ & $z_{\mathrm{SB3}}$ & $z_{\mathrm{SB4}}$ & t$_{\mathrm{exp}}$ & Noise R.M.S. \\
   &      &                    &                    &                    &                    & (min)              &    (mJy/beam) \\
(1) & (2) & (3) & (4) & (5) & (6) & (7) & (8) \\
\midrule
9347 & 03 Nov 2013 & 1.8388 -- 1.8468 & 1.8468 -- 1.8548 & 1.9014 -- 1.9098 & 1.9158 -- 1.9242 & 17.14 & 16.83 \\
6515 & 03 Nov 2013 & 1.8388 -- 1.8468 & 1.8468 -- 1.8548 & 1.9014 -- 1.9098 & 1.9158 -- 1.9242 & 17.14 & 15.76 \\
10076& 04 Nov 2013 & 1.8771 -- 1.8852 & 1.8852 -- 1.8935 & 1.9332 -- 1.9418 & 1.9418 -- 1.9503 & 10.58 & 21.69 \\
\textbf{9834} & \textbf{04 Nov 2013} & \textbf{1.7518 -- 1.7593} & \textbf{1.7593 -- 1.7668} & \textbf{1.7668 -- 1.7744} & \textbf{1.7744 -- 1.7820} & \textbf{11.09} & \textbf{15.39} \\
9681 & 04 Nov 2013 & 1.8771 -- 1.8852 & 1.8852 -- 1.8935 & 1.9332 -- 1.9418 & 1.9418 -- 1.9503 & 10.58 & 18.72 \\
10049& 03 Nov 2013 & 1.8388 -- 1.8468 & 1.8468 -- 1.8548 & 1.9014 -- 1.9098 & 1.9158 -- 1.9242 & 15.12 & 11.60 \\
2861 & 04 Nov 2013 & 1.7213 -- 1.7291 & 1.7291 -- 1.7364 & 1.8024 -- 1.8102 & 1.8102 -- 1.8180 & 9.58 & 30.10 \\ 
\textbf{2910} & \textbf{04 Nov 2013} & \textbf{1.7518 -- 1.7593} & \textbf{1.7593 -- 1.7668} & \textbf{1.7668 -- 1.7744} & \textbf{1.7744 -- 1.7820} & \textbf{11.09} & \textbf{14.36} \\
7118 & 04 Nov 2013 & 1.7213 -- 1.7291 & 1.7291 -- 1.7364 & 1.8024 -- 1.8102 & 1.8102 -- 1.8180 & 9.58 & 51.00 \\
8490 & 03 Nov 2013 & 1.8388 -- 1.8468 & 1.8468 -- 1.8548 & 1.9014 -- 1.9098 & 1.9158 -- 1.9242 & 16.13 & 15.44 \\
\bottomrule
\end{tabular}
\begin{minipage}{18cm}
\smallskip
\footnotesize
\textbf{Columns} (1) Galaxy ID; (2) Date of observations; (3) Redshift range covered by the ALMA sideband \#1; (4) Redshift range covered by the ALMA sideband \#2; (5) Redshift range covered by the ALMA sideband \#3; (6) Redshift range covered by the ALMA sideband \#4. For the sources highlighted in bold all the four sidebands are contiguous; (7) Integration time on source; (8) Noise r. m. s.
\end{minipage}
\end{table*}

However, the returned fluxes critically depend on the model size that we used and that we determined from the optical images. If we were to use a smaller (larger) size, the inferred flux would be correspondingly lower (higher). Unfortunately, the size of the emission cannot be constrained from the data on individual sources, given the limited S/N ratio.
There have been claims that
sizes estimated from optical data could be larger than those
derived from IR observations (\citealt{Diaz-Santos2013}, \citealt{Psychogyios2016}). This could
possibly bias our analysis and in particular our flux estimates to higher values. As a check, we aligned our \cii~ detections at the \textit{HST} positions and stacked them (coadding all visibilities) to increase
the S/N (Figure \ref{fig:ampl}). In the \textit{uv} space the overall
significance of the stacked detection is $\sim 10\sigma$. The probability that the signal is not resolved (i.e., a point source, which would have constant amplitude versus \textit{uv} distance) is $<10^{-5}$.
We then fitted the stacked data with an exponential profile, leaving its
size free to vary during the fit. We get an exponential scale length for the \cii~ emission of 
$0.65\pm0.15''$ (corresponding to $\sim$ 4 -- 7 kpc), corrected for the small broadening that could affect
the stack due to the uncertainties in the determination of the
sources' exact position, and with a significance of S/N(size)
$\sim 4\sigma$. 
The reported size uncertainty was estimated by GILDAS
in the fit and the modelling of the signal amplitude versus
\textit{uv} range signal shows that it is reliable (Figure \ref{fig:ampl}). This
indicates that on average the optical sizes that we used in the
analysis are appropriate for the fit of our ALMA data and that these
four galaxies are indeed quite extended (the average optical size of
the 4 galaxies is $\sim$ 0.7$''$, $2\sigma$, in good agreement with what measured in the \cii\ stack).

We also used the stack of our four detected sources to further check
our \cii\ flux estimates. We compared the flux measured by fitting the stacking
with that obtained by averaging the fluxes of individual
detections. As mentioned above, the flux of the stacking critically depends on the adopted
model size, but in any case the measurement was highly significant (S/N$>5$) even when leaving
the size free to vary during the fit. When fitting the
stack with a model having an exponential scale length $\sim 0.6$'' we
obtained estimates consistent with the average flux of individual
sources.

\subsubsection{Tentative and non detections}
\label{subsec:tentative_det}

In our sample, six sources were not individually detected by the
procedure discussed in the previous section. In these cases we
searched for the presence of weaker \cii\ signal in the data by
evaluating the recovered signal when eliminating all degrees of
freedom in the line search, namely measuring at fixed \textit{HST} position, using exponential models with the fixed optical size for each galaxy
and conservatively averaging the signal over a large velocity range
tailored to the optical redshifts. In particular, we created emission
line maps by averaging channels over 719 km s$^{-1}$, around the
frequency corresponding to the optical redshift. This velocity width
is obtained by summing in quadrature 3 times the MAD redshift accuracy
(obtained considering optical and \cii\ redshifts, as discussed  above
for the four detections) and the average FWHM of the detected emission
lines. We find weak signal from two galaxies
at S/N$>2.3$ (ID9681 and ID8490, see Table \ref{tab:properties}) and no significant
signal from the others. 
Given that with this approach there are no degrees of freedom, the probability of obtaining each tentative detection (namely the probability of having a $> 2.3 \sigma$ signal) is  Gaussian and equal to $\sim$ 0.01.
Furthermore, when considering the six sources discussed above, we expect to find $<0.1$ false detections. We therefore conclude that the $2.3\sigma$ signal found for our two tentative detections is real.

For the four sources with no detected signal we considered $3\sigma$ flux upper limits, as estimated from emission line maps integrated over a 719 km s$^{-1}$ bandwidth.
There are different possible reasons why these galaxies do not show any signal. Two of them (ID7118 and ID2861) have substantially worse data quality, probably due to the weather stability and atmosphere transparency during the observations, with about 3 times higher noise than the rest of the sample. Their $L_{\mathrm{\cii}}/L_{\mathrm{IR}}$ upper limits are not very stringent and are substantially higher than the rest of the sample (Table \ref{tab:properties}).
Possible reasons for the other two non detections
(ID2910 and ID10049) are the following. (i) These sources might be more extended than the others, and therefore their signal might be further suppressed. However this is unlikely, as their optical size is smaller than the average one of the detected sources (Table \ref{tab:measurements}). (ii) They might have fainter IR luminosity than the other sample galaxies. The 
$L_{\mathrm{IR}}$ that we used to predict the \cii~ luminosity for these
two undetected sources was overestimated before the observations. However, using the current $L_{\mathrm{IR}}$ values (Section \ref{subsec:continuum_measurements}), we obtain $L_{\mathrm{\cii}}/L_{\mathrm{IR}}$ upper limits comparable with the ratios estimated for the detected sources. (iii) A wrong optical redshift estimate can also explain the lack of signal from one of these undetected galaxies: ID10049 is an AGN with broad lines\footnote{We recall that all our IR luminosities are estimated considering the star forming component only, and possible emissions from dusty tori were subtracted.}, and the determination of its systemic redshift obtained considering narrow line components ($z=1.920$) is possibly more uncertain than the redshift range covered by our ALMA observations ($z=1.9014$--1.9098 and $z=1.9158$--1.9242; Table \ref{tab:log}; for comparison, the original literature redshift was 1.906). For ID2910 instead the optical spectrum seems to yield a solid redshift and the covered redshift range is the largest (Table \ref{tab:log}), so the \cii\ line should have been observed. This source probably has fainter \cii~ luminosity than the others (i.e. lower $L_{\mathrm{\cii}}/L_{\mathrm{IR}}$).

Finally, we stacked the four \cii~ non-detections in the $uv$ plane
and fitted the data with an exponential profile, with size fixed to
the average optical size of the sources entering the stacking. This
still did not yield a detection. Since two non-detections have shallower data than the others and at least one might have wrong optical redshift, in the rest of the analysis we do not consider the average \cii~ flux obtained from the stacking of these sources.

\vspace{0.5cm}
The coordinates, sizes, \cii~ fluxes, and luminosities of our sample galaxies are
presented in Table \ref{tab:properties}. We subtracted from the \cii~ fluxes the contribution of the underlying 158 $\mu$m
rest-frame continuum as measured in our ALMA Band 9 data (Section \ref{subsec:continuum_measurements}). For galaxies with no detected continuum at 450 $\mu$m
(see Section \ref{subsec:continuum_measurements}), we
computed the predicted 158~$\mu$m rest-frame continuum flux from the
best-fit IR SEDs and reduced the \cii~ fluxes accordingly.

\subsubsection{Average \cii\ signal}
\label{subsec:average_cii}
We have previously stacked the four detections to measure their
average size, compare it with the optical one, and understand if we
were reliably estimating the fluxes of our sources (Section
\ref{subsec:cii_detections}). Now we want to estimate the average
\cii~ signal of our sample to investigate its mean behaviour. We
therefore add to the previous stack also the two tentative
detections and one non-detected source. We report in the following the
method that we used to stack these galaxies and the reasons why we excluded
from the stack three non-detected sources.

We aligned the detections and tentative detections and stacked them
coadding all visibilities. We also coadded the non-detected galaxy
ID2910, but we do not include the other three sources for reasons outlined above. 
. We fitted the resulting map with an exponential model with
size fixed to the average optical size of the sources entering the
stacking. We finally subtracted the contribution of the rest-frame 158
$\mu$m continuum by decreasing the estimated flux by 10\% (namely the
average continuum correction applied to the sources of our sample, see
Section \ref{subsec:continuum_measurements}). We obtained a $\sim
10\sigma$ detection that we report in Table \ref{tab:properties}. 

\begin{landscape}
\begin{table}
\centering
\caption{Measurements for our sample galaxies}
\label{tab:properties}
\begin{tabular}{c c c c c c c c c c c c}
\toprule
\midrule
ID & RA & DEC & $z_\mathrm{opt}$ & $z_\mathrm{\cii}$ &  $F_{\mathrm{450\mu m}}$ & $F_{\mathrm{850\mu m}}$ & $F_{\mathrm{\cii}}$ & $L_{\mathrm{\cii}}$ & log($L_{\mathrm{IR}}$) & $L_{\mathrm{\cii}}/L_{\mathrm{IR}}$ & $\Delta v$ \\[3pt]
 & [deg] & [deg]  & & & [mJy] & [mJy] & [mJy] & [10$^{9}$ L$_{\odot}$] & [log(L$_{\odot}$)] & [$10^{-3}$] & [km s$^{-1}$] \\[3pt]
(1) & (2) & (3) & (4) & (5) & (6) & (7) & (8) & (9) & (10) & (11) & (12) \\ [3pt]
\hline
\rule{0pt}{4ex}
9347   & 53.154900 & -27.809397 & 1.8503 $\pm$ 0.0010 & 1.8505 $\pm$ 0.0002 & $<$ 8.85 & 0.75 $\pm$ 0.24 & 21.28 $\pm$ 6.73 & 0.95 $\pm$ 0.30 & 11.80 $\pm$ 0.05 &  $1.51^{+0.51}_{-0.50}$  & 534.3 \\
6515   & 53.073375 & -27.764353 & 1.8440 $\pm$ 0.0010 & 1.8438 $\pm$ 0.0002 & $<$5.76  & 0.71 $\pm$ 0.18 & 24.50 $\pm$ 6.57 & 1.23  $\pm$ 0.33 & 11.68 $\pm$ 0.04 & $2.57^{+0.73}_{-0.73}$ & 365.4 \\
10076 & 53.045904 & -27.822156 & 1.9418 $\pm$ 0.0020 & 1.9462 $\pm$ 0.0006 & $<$9.69 & $<$0.57 & 29.03 $\pm$ 9.14  & 2.40   $\pm$ 0.76 & 11.91 $\pm$ 0.03  & $2.95^{+0.96}_{-0.96}$  & 548.1 \\
9834   & 53.181029 & -27.817147 & 1.7650 $\pm$ 0.0020 & 1.7644 $\pm$ 0.0003 & $<$4.52 & $<$0.45 & 15.34 $\pm$ 2.21 & 1.29   $\pm$ 0.19 & 11.99 $\pm$ 0.02 & $1.32^{+0.22}_{-0.22}$  & 627.3 \\
9681   & 53.131350 & -27.814922 & 1.8852 $\pm$ 0.0010 & -- &$<$8.04 & 1.01 $\pm$ 0.24 & 17.59 $\pm$ 7.63 & 1.81 $\pm$ 0.79& 11.84 $\pm$ 0.04 & $2.62^{+1.17}_{-1.16}$ & 719.0 \\
10049 & 53.180149 & -27.820603 & 1.9200$^a$& -- & $<$4.32  & 0.77 $\pm$ 0.16 & $<5.65$  & $<$0.60    & 11.60 $\pm$ 0.06 & $<$1.51  & 719.0  \\
2861   & 53.157905 & -27.704283 & 1.8102 $\pm$ 0.0010 & -- & $<$15.35    & 1.56 $\pm$ 0.28 & $<$40.11     & $<$3.84       & 12.00 $\pm$ 0.03 & $<$3.84  & 719.0 \\
2910   & 53.163610 & -27.705320 & 1.7686 $\pm$ 0.0010 & -- & $<$5.94                   & $<$0.54               & $<$12.73 & $<$1.17    & 11.76 $\pm$ 0.08 & $<$2.03 & 719.0 \\
7118   & 53.078130 & -27.774187 & 1.7290$^a$& -- &$<$16.5                  & 1.05 $\pm$ 0.29          & $<$56.16   & $<$4.94   & 12.06 $\pm$ 0.01 & $<$4.30 & 719.0 \\
8490   & 53.140593 & -27.795632 & 1.9056 $\pm$ 0.0010 & -- &$<$4.5                      & $<$0.48               & 6.80 $\pm$ 2.85 & 0.71 $\pm$ 0.30  & 11.54 $\pm$ 0.06 & $2.05^{+0.92}_{-0.90}$  & 719.0 \\
Stack$^b$& --       &  --                & 1.8536 $\pm$ 0.004 & -- &       --     & --     & 15.59 $\pm$ 1.79        & 1.25 $\pm$ 0.14   & 11.81 $\pm$ 0.05 &  $1.94^{+0.34}_{-0.32}$  & 604.6 \\           
\bottomrule
\end{tabular}
\begin{minipage}{22cm}
\smallskip
\footnotesize
\textbf{Columns} (1) Galaxy ID; (2) Right ascension; (3) Declination;
(4) Redshift obtained from optical spectra; (5) Redshift estimated by fitting the
\cii~ emission line (when detected) with a
Gaussian in our 1D ALMA spectra. The uncertainty that we report is the
formal error obtained from the fit; (6) Observed-frame 450$\mu$m
continuum emission flux; (7) Observed-frame 850$\mu$m continuum flux;
(8) \cii~ emission line flux. We report upper limits for sources with
S/N $< 2$; (9) \cii~ emission line luminosity;  (10) IR luminosity integrated over the
wavelength range 8 -- 1000 $\mu$m as estimated from SED fitting (Section \ref{subsec:continuum_measurements}); (11) \cii -to-bolometric infrared luminosity ratio; (12) Line velocity width. \\
\textbf{Notes} $^a$ID10049 is a broad line AGN, its systemic redshift
is uncertain and it might be outside the frequency range covered by
Band 9. The redshift of ID7118 is based on a single line identified as
H$\alpha$. If this is correct the redshift uncertainty is $<0.001$. 
$^b$ Stack of the 7 galaxies of our sample with reliable \cii~ measurement (namely, ID9347, ID6515, ID10076, ID9834, ID9681, ID8490, ID2910, see Section \ref{subsec:average_cii} for a detailed discussion). We excluded from the stack ID2861 and ID7118 since the quality of their data is worse than for the other galaxies and their \cii~ upper limits are not stringent. We also excluded ID10049 since it is an AGN and, given that its redshift estimate from optical spectra is highly uncertain, the \cii~ emission might be outside the redshift range covered by our ALMA observations. See Section \ref{subsec:tentative_det} for a detailed discussion.
\end{minipage}
\end{table}
\end{landscape}

The average $L_{\mathrm{\cii}}/L_{\mathrm{IR}}$ ratio obtained
  from the stacking of the seven targets mentioned above is
  $(1.94^{+0.34}_{-0.32})\times 10^{-3}$. This is in agreement with
  that obtained by averaging the individual ratios of the same seven
  galaxies ($L_{\mathrm{\cii}}/L_{\mathrm{IR}} =
  (1.96^{+0.19}_{-0.10})\times 10^{-3}$) where this ratio was obtained
  averaging the $L_{\mathrm{\cii}}/L_{\mathrm{IR}}$ ratio of the seven
  targets. In particular, the \cii~ flux of ID2910 is an upper limit
  and therefore we considered the case of flux equal to 1$\sigma$
  (giving the average $L_{\mathrm{\cii}}/L_{\mathrm{IR}}$ =
  $1.96\times 10^{-3}$) and the two extreme cases of flux equal to 0
  or flux equal to 3$\sigma$, from where the quoted
  uncertainties. Through our analysis and in the plots we consider the
  value $L_{\mathrm{\cii}}/L_{\mathrm{IR}}=
  (1.94^{+0.34}_{-0.32})\times 10^{-3}$.

\begin{table*}
\center
\caption{Physical properties of our sample galaxies}
\label{tab:measurements}
\begin{tabular}{c c c c c c c c c c c}
\toprule
\midrule
ID & SFR & log(M$_{\star}$) & log M$_{\mathrm{dust}}$ & log M$_{\mathrm{mol}}^{\mathrm{SK}}$ & log M$_{\mathrm{mol}}^{\mathrm{dust}}$ & log M$_{\mathrm{mol}}^{\mathrm{[CII]}}$  & sSFR/sSFR$_{\mathrm{MS}}$ & $\log <$U$>$ & R$_{\mathrm{e}}$ & Z \\[3pt]
 & [M$_\odot$ yr$^{-1}$] & [log(M$_\odot$)] & [log(M$_\odot$)] &  [log(M$_\odot$)] & [log(M$_\odot$)] & [log(M$_\odot$)] & & & [arcsec] & \\[3pt]
(1) & (2) & (3) & (4) & (5) & (6) & (7) & (8) & (9) & (10) & (11) \\[3pt]
\midrule
9347 & 62.9$^{+7.9}_{-7.0}$   & 10.5  &  8.5$\pm$0.5 & 10.70 & 10.50$\pm$0.57 & 10.51$^{+0.13}_{-0.19}$ & 1.1 & 1.2$\pm$0.5 & 1.02 & 8.6 \\
6515   & 47.7$^{+4.3}_{-3.9}$ & 10.9  & 8.5$\pm$0.4 & 10.58  & 10.40$\pm$0.42 & 10.62$^{+0.12}_{-0.16}$ & 0.4 &  1.2$\pm$0.4 & 0.77 & 8.7 \\
10076 & 81.6$^{+6.0}_{-5.6}$ & 10.3  & 8.4$\pm$0.2 & 10.77 & 10.46$\pm$0.23 & 10.91$^{+0.13}_{-0.19}$ & 1.7 & 1.4$\pm$0.2 & 0.76 & 8.6 \\
9834   & 98.9$^{+5.4}_{-5.1}$ & 10.7  & 8.2$\pm$0.3 & 10.84 & 10.20$\pm$0.16 & 10.60$^{+0.10}_{-0.12}$ & 1.2 & 1.7$\pm$0.3 & 0.43 & 8.7 \\
9681  & 69.3$^{+5.9}_{-5.5}$ &10.6  & 8.3$\pm$0.5 & 10.71 & 10.29$\pm$0.49 & 10.78$^{+0.16}_{-0.27}$ & 1.0 & 1.5$\pm$0.5 & 0.89 & 8.6 \\
10049 & 39.7$^{+5.7}_{5.0}$ & 10.7  & 8.7$\pm$0.2 & 10.52 & 10.70$\pm$0.29 & $< 10.37$ & 0.4 & 0.8$\pm$0.2 & 0.29 & 8.7 \\
2861   & 101.6$^{+6.5}_{-6.1}$ & 10.8  & 9.0$\pm$0.3 &10.85 & 10.97$\pm$0.30 & $< 11.13$ & 1.1 & 0.9$\pm$0.3 & 0.99 & 8.7 \\
2910   & 57.4$^{+11.1}_{-9.3}$ & 10.4  & 8.1$\pm$0.5 & 10.64 & 10.18$\pm$0.55 & $< 10.59$ & 1.3 & 1.5$\pm$0.5 & 0.58 & 8.6 \\
7118   & 114.8$^{+2.9}_{-2.9}$ & 10.9  & 9.1$\pm$0.2 & 10.89 & 11.03$\pm$0.22 & $< 11.21$ & 1.1 & 0.9$\pm$0.2 & 1.13 & 8.7 \\
8490  & 34.4$^{+5.2}_{-4.5}$ & 10.0  & 7.8$\pm$0.4 & 10.46 & 9.98$\pm$0.45 & 10.38$^{+0.16}_{-0.26}$ & 1.2 & 1.6$\pm$0.4 & 0.44 & 8.5 \\
Stack$^a$& 64.6$^{+7.9}_{-7.0}$ & 10.6&8.3$\pm$0.1 & 10.69 & 10.26$\pm$0.34 & 10.62$^{+0.04}_{-0.05}$ & 1.1 & 1.4$\pm$0.1 & 0.70 & 8.6 \\
\bottomrule
\end{tabular}
\begin{minipage}{18cm}
\smallskip
\footnotesize
\textbf{Columns} (1) Galaxy ID; (2) Star formation rate as calculated
from the IR luminosity: $SFR = 10^{-10}L_{\mathrm{IR}}$
\citep{Kennicutt1998}. Only the star-forming component contributing
to the IR luminosity was used to estimate the SFR, as contribution from a dusty
torus was subtracted; (3) Stellar mass. The typical uncertainty is $\sim$ 0.2 dex; (4) Dust mass; (5)
Gas mass estimated from the integrated Schmidt-Kennicutt relation
\citep[Equation 4]{Sargent2014}. The measured dispersion of the
relation is 0.2 dex. Given that the errors associated to the
SFR are $< 0.1$ dex, for the M$_{\mathrm{mol}}^{\mathrm{SK}}$ we consider typical uncertainties of 0.2 dex.; (6) Gas mass estimated from the dust mass considering a gas-to-dust conversion factor dependent on metallicity
\citep{Magdis2012}; (7) Gas mass estimated from the observed \cii~
luminosity considering a \cii -to-H$_2$ conversion factor
$\alpha_{\mathrm{\cii}} = 31$ M$_\odot$/L$_\odot$. The uncertainties
that we report do not account for the $\alpha_{\rm \cii}$
uncertainty and they only reflect the \cii~ luminosity's uncertainty; (8) Distance from the main sequence as defined by \cite{Rodighiero2014}; (9) Average
radiation field intensity; (10) Galaxy size as measured from the
optical \textit{HST} images; (11) Gas-phase metallicity $12 + \log(\mathrm{O/H})$.\\ 
\textbf{Notes} $^a$ Stack of the 7 galaxies of our sample with reliable \cii~ measurement (namely, ID9347, ID6515, ID10076, ID9834, ID9681, ID8490, ID2910). We excluded from the stack ID2861 and ID7118 since the quality of their data is worse than for the other galaxies and their \cii~ upper limits are not stringent. We also excluded ID10049 since it is an AGN and, given that its redshift estimate from optical spectra is highly incertain, the \cii~ emission might be outside the redshift range covered by our ALMA observations. See Section \ref{subsec:tentative_det} for a detailed discussion.
\end{minipage}
\end{table*}

\subsection{Continuum emission at observed-frame 450 $\mu$m and 850 $\mu$m}
\label{subsec:continuum_measurements}

Our ALMA observations cover the continuum at $\sim$ 450 $\mu$m (Band 9 data)
and 850 $\mu$m (Band 7 data). We created averaged continuum maps
by integrating the full spectral range for the observations at 850
$\mu$m. For the 450 $\mu$m continuum maps instead we made sure to
exclude the channels where the flux is dominated by the \cii~ emission line.

We extracted the continuum flux by fitting the data with an
exponential profile, adopting the same procedure described in Section
\ref{subsec:cii_measurements}. The results are provided in Table
\ref{tab:properties}, where 3$\sigma$ upper limits are reported in
case of non-detection.

The estimated continuum fluxes were used, together with the
available \textit{Spitzer} and \textit{Herschel} data \citep{Elbaz2011}, to properly sample the IR
wavelengths, perform SED fitting, and reliably determine
parameters such as the infrared luminosity and the dust mass
($M_{\mathrm{dust}}$). The \textit{Spitzer} and \textit{Herschel} data were deblended using prior sources to overcome the blending problems arising from the large PSFs and allow reliable photometry of individual galaxies (\citealt{Bethermin2010}, \citealt{Roseboom2010}, \citealt{Elbaz2011}, \citealt{Lee2013}, \citealt{Bethermin2015}, \citealt{Liu2017}).
Following the method presented in \cite{Magdis2012}, we fitted the IR
photometry with \cite{Draine2007} models, supplemented by the use of a
single temperature modified black body (MBB) fit to derive a
representative dust temperature of the ISM. In these fits we
considered the measured \textit{Spitzer}, \textit{Herschel}, and ALMA
flux (even if S/N $<$ 3, e.g. there is no detection) along with the
corresponding uncertainty instead of adopting upper limits. The
contribution of each photometric point to the best fit is weighted by
its associated uncertainty. If we were to use upper limits in these
fits instead our conclusions would not have changed. The IR SEDs of our targets are shown in Figure \ref{fig:sed} and the derived
parameters are summarized in Table \ref{tab:measurements}. We note that our method to estimate dust masses is based on the fit of the full far-IR SED of the galaxies, not on scaling a single band luminosity in the Rayleigh-Jeans regime (e.g. as suggested by \citealt{Scoville2017}). This fact together with the high quality photometry at shorter
wavelengths allowed us to properly constrain the fitted parameters also for galaxies with highly uncertain 850 $\mu$m measurements.
We also determined the average radiation field
intensity as $\langle U\rangle = L_{\mathrm{IR}}/(125 M_{\mathrm{dust}})$
\citep{Magdis2012}. Uncertainties on $L_{\mathrm{IR}}$ and $M_{\mathrm{dust}}$ were quantified using Monte Carlo simulations, as described by \cite{Magdis2012}. 

The IR luminosities we estimated ($L_{\mathrm{IR}} = L[8 $--$ 1000\,
\mu$m]) for our sample galaxies lie between $3.5 \times 10^{11}$ --
$1.2 \times 10^{12}$ L$_{\odot}$, with a median value of $7.1 \times
10^{11}$ L$_{\odot}$, and we probe a range of dust masses between $7.0
\times 10^7$ -- $1.2 \times 10^9$ M$_{\odot}$, with a median value
of $3.0 \times 10^8$ M$_{\odot}$. Both our median estimate of
$L_{\mathrm{IR}}$ and $M_{\mathrm{dust}}$ are in excellent agreement
with literature estimates for main-sequence galaxies at similar redshift
(e.g. $L_{\mathrm{IR}} =
6\times 10^{11}$ L$_\odot$ and $M_{\mathrm{dust}} = 3\times 10^8$
M$_\odot$ at redshift $1.75 < z < 2.00$ in \citealt{Bethermin2015}, for a mass selected sample
with an average $M_\star$ comparable to that of our galaxies).
The $\langle U\rangle$ parameters that we determined range between 6 -- 45, consistent with the estimates provided by
\citet{Magdis2012} and \citet{Bethermin2015} for main-sequence galaxies at a similar redshift.

Finally, we estimated the molecular gas masses of our galaxies with a twofold approach. (1) Given their stellar mass and the
mass-metallicity relation by \cite{Zahid2014} we estimated their gas phase
metallicity. We then determined the gas-to-dust
conversion factor ($\delta_{\mathrm{GDR}}$) for each source, depending
on its metallicity, as prescribed by \cite{Magdis2012}. And finally we
estimated their molecular gas masses as $M_{\mathrm{mol}} = \delta_{\mathrm{GDR}}\times
M_{\mathrm{dust}}$, given the dust masses obtained from the SED fitting. (2) Given the galaxies SFRs and the integrated Schmidt-Kennicutt relation for main-sequence sources reported by \cite{Sargent2014}, we estimated their molecular gas masses. We estimated the uncertainties taking into account the SFR uncertainties and the dispersion of the Schmidt-Kennicutt relation. By comparing the galaxies detected in the ALMA 850 $\mu$m data that allow us to obtain accurate dust masses, we concluded that both methods give consistent results (see Table \ref{tab:measurements}). In the following we use the $M_{\mathrm{mol}}$ obtained from the Schmit-Kennicutt relation since, given our in-hand data, it is more robust especially for galaxies with no 850 $\mu$m detection. Furthermore, it allows us to get a more consistent comparison with other high-$z$ literature measurements (e.g. the gas masses for the sample of \citealt{Capak2015} have been derived using the same Schmidt-Kennicutt relation, as reported in Appendix \ref{app:literature}).

\subsection{Other samples from the literature}
\label{subsec:literature}
To explore a larger parameter space and gain a more comprehensive
view, we complemented our observations with multiple \cii~ datasets from the
literature, both at low and high redshift (\citealt{Stacey1991}, \citealt{Stacey2010},
\citealt{Gullberg2015}, \citealt{Capak2015},
\citealt{Diaz-Santos2017}, \citealt{Cormier2015},
\citealt{Brauher2008}, \citealt{Contursi2017}, \citealt{Magdis2014},
\citealt{Huynh2014}, \citealt{Ferkinhoff2014}, \citealt{Schaerer2015}, \citealt{Brisbin2015}, \citealt{Hughes2017}, \citealt{Accurso2017}). In Appendix
\ref{app:literature} we briefly present these additional samples and
discuss how the physical parameters that are relevant for our analysis (namely the redshift,  \cii, IR, and CO luminosity, molecular gas mass, sSFR, and gas-phase metallicity) have been derived; in Table \ref{tab:acii} we report them.

\section{Results and Discussion}
\label{sec:results}

The main motivation of this work is to understand which is the dominant physical parameter affecting the \cii\ luminosity of galaxies through cosmic time. In the following we investigate whether our $z\sim 2$ sources are \cii~ deficient and if the \cii -to-IR luminosity ratio depends on galaxies' distance from the main-sequence. We also investigate whether the \cii~ emission can be used as molecular gas mass tracer for main-sequence and starburst galaxies both at low and high redshift. Finally we discuss the implications of our results on the interpretation and planning of $z \gtrsim 5$ observations. 

\subsection{The \cii~ deficit}
\label{subsec:cii_deficit}

\begin{figure*}
\centering
\includegraphics[width=0.7\textwidth]{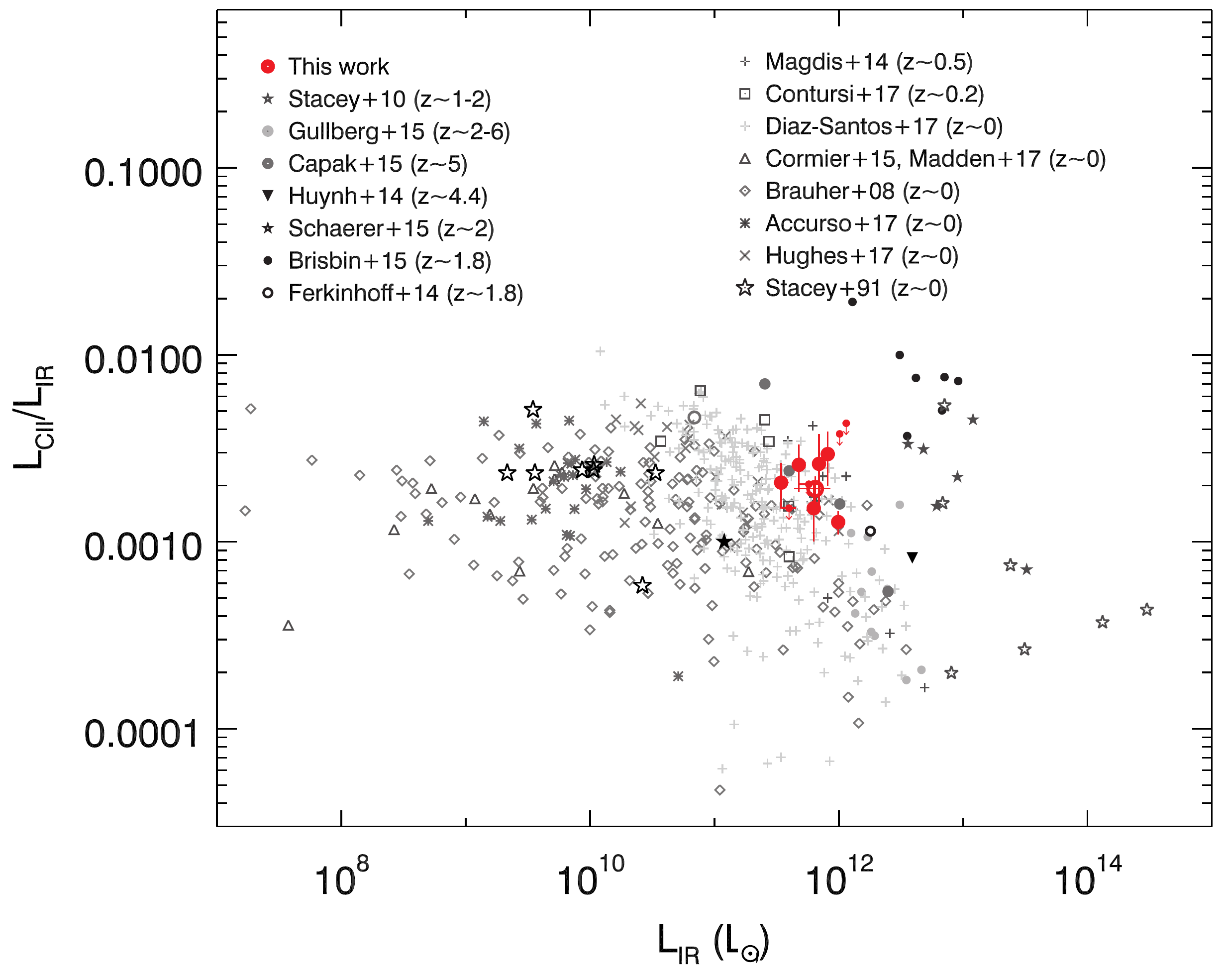}
\caption{Ratio between the \cii~ and IR (8 -- 1000 $\mu$m) luminosity
  of our sample galaxies, as a function of the IR
  luminosity. Different symbols indicate distinct datasets: our \cii~
  detections (large red filled circles), our \cii~ upper limits for
  the non-detections (small red filled circles), the average value of our sample (empty red circle), \citet[grey filled stars
  indicate star-forming galaxies, grey empty stars indicate AGN or
  starbursts]{Stacey2010}, \citet[black empty stars]{Stacey1991}, \citet[light grey filled
  circles]{Gullberg2015}, \citet[dark grey filled circles indicate
  their measurements, dark grey emtpy circles indicate the stack of
  their non detections]{Capak2015}, \citet[grey
  crosses]{Diaz-Santos2017}, \citet[black triangles]{Cormier2015},
  \citet[grey diamond]{Brauher2008}, \citet[grey
  squares]{Contursi2017}, \citet[dark grey crosses]{Magdis2014},
  \citet[black downward triangle]{Huynh2014}, \citet[black filled
  star]{Schaerer2015}, \citet[black filled circles]{Brisbin2015},
  \citet[black empty circle]{Ferkinhoff2014}, \citet[grey
  crosses]{Hughes2017}, \citet[grey asterisks]{Accurso2017}. We note
  that we are plotting the de-magnified IR luminosity for the sample
  of lensed galaxies by \citet{Gullberg2015}: we considered that the
  \cii~ emission line is magnified by the same factor as the IR (see
  discussion in the text and \citealt{Gullberg2015}). The
  magnification factors are taken from \citet{Spilker2016}. Similarly,
  the sources by \citet{Brisbin2015} might be lensed, but the
  magnification factors are unknown and therefore we plot the observed
  values.}
\label{fig:lcii_lir}
\end{figure*}

\begin{figure}
\centering
\includegraphics[width=0.5\textwidth]{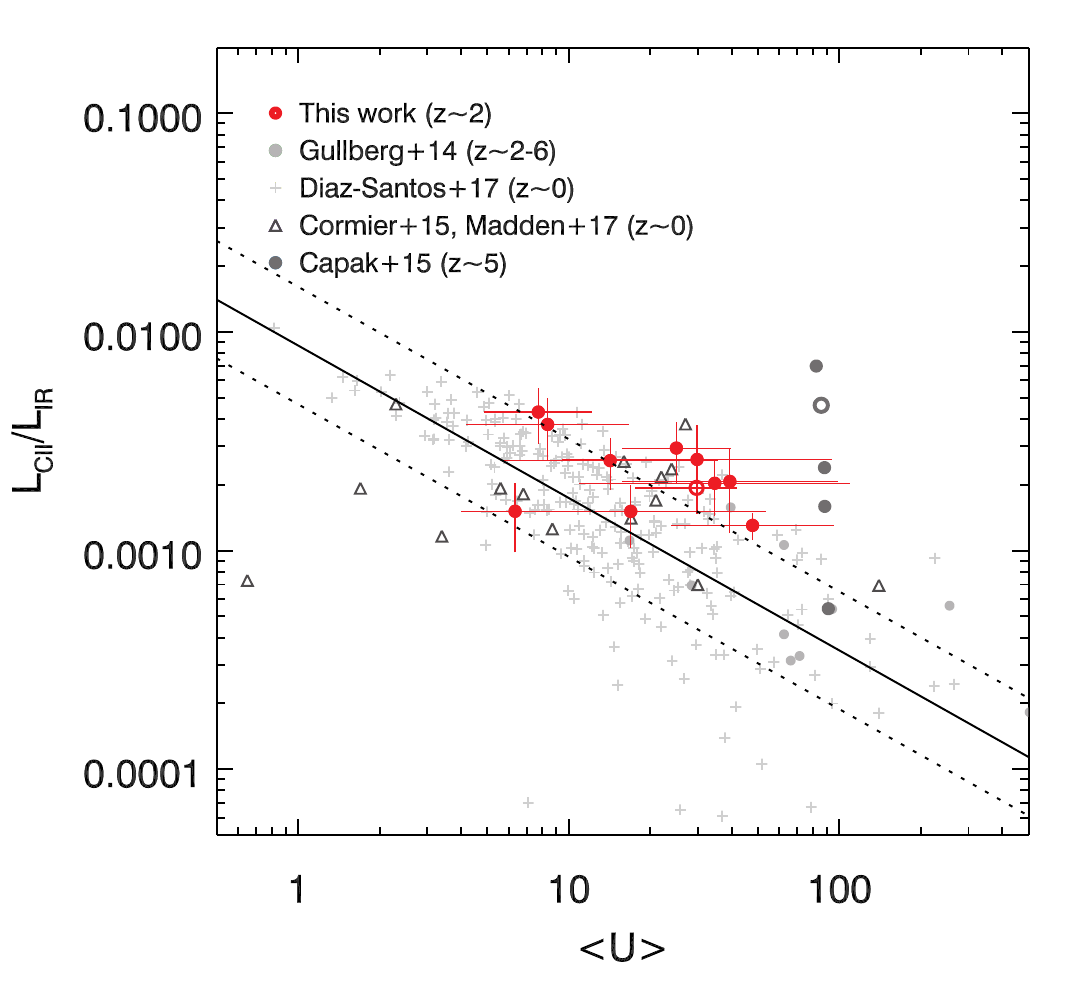}
\caption{Correlation between the \cii -to-IR luminosity ratio and the intensity of the radiation field. The symbols are the same as reported in Figure \ref{fig:lcii_lir} caption, but we only show the samples with available $\langle U\rangle$ measurements (the method used to estimate $\langle U\rangle$ for the various samples is detailed in Section \ref{subsec:cii_deficit}). The fit of the local sample from \citet{Diaz-Santos2017} is reported (black solid line) together with the standard deviation (black dashed lines).} 
\label{fig:u}
\end{figure}

\begin{figure*}
\centering
\includegraphics[width=\textwidth]{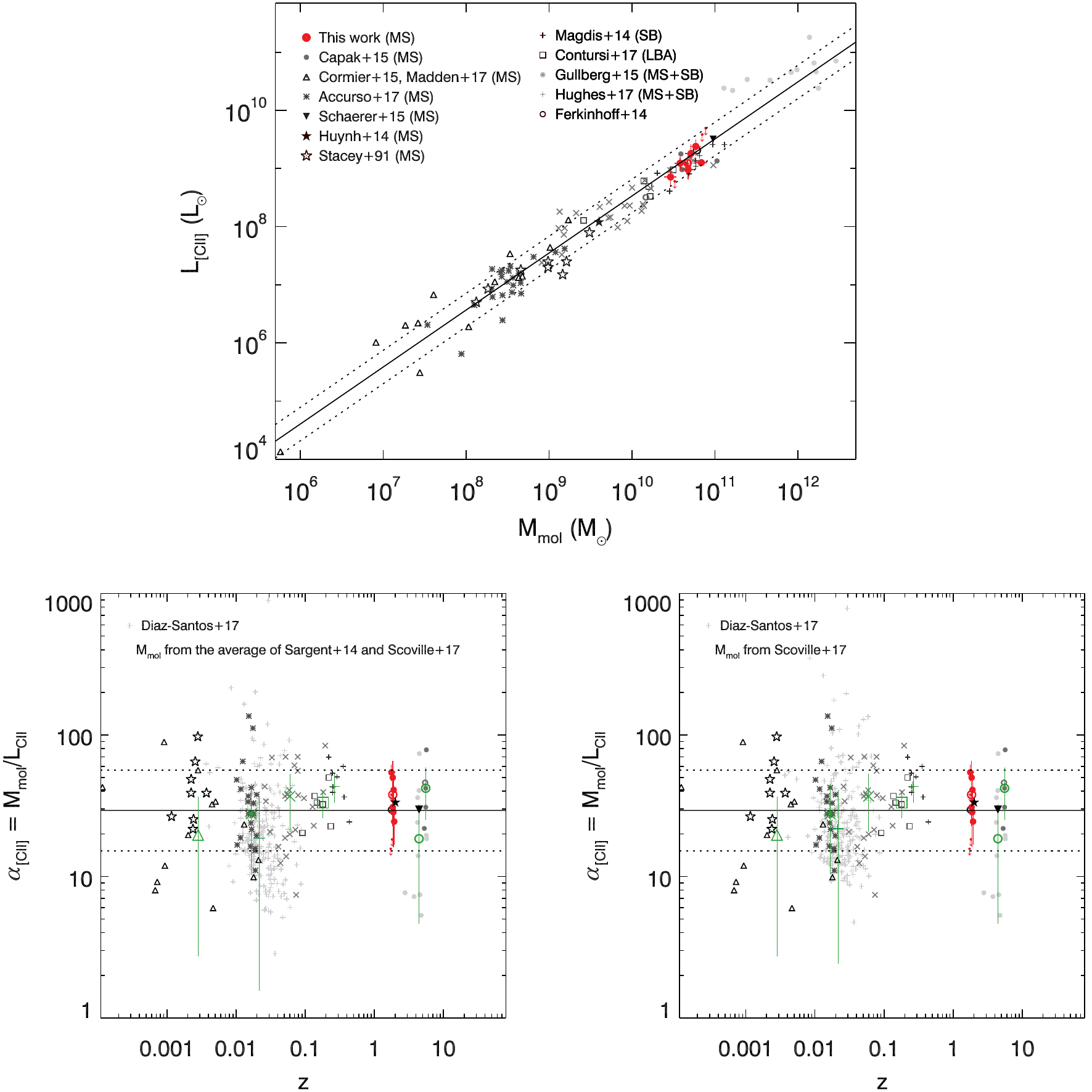}
\caption{Correlation between the \cii~ luminosity and the molecular gas
  mass. Top panel: L$_{\mathrm{\cii}}$ -- M$_{\mathrm{mol}}$
  relation. The
  symbols are the same as reported in Figure \ref{fig:lcii_lir}
  caption, but we only show the samples with available
  M$_{\mathrm{mol}}$ estimates. In the legend we highlight the nature
  of the galaxies in each sample (e.g. main-sequence, starburst, Ly
  break analogs). The fit of the data is reported (black
  solid line) together with the standard deviation (black dashed
  lines). Bottom panels: the \cii -to-H$_2$ conversion factor ($\alpha_{\mathrm{\cii}}$)
  as a function of redshift. The average $\alpha_{\mathrm{\cii}}$ for main-sequence galaxies is
reported (black solid line) together with the standard deviation (black dashed
lines). The median and median absolute deviation of each sample is
plotted (green large symbols). The difference between the left and
right panels concerns how the molecular gas mass was estimated for the
sample of local galaxies from \citet[light gray
crosses]{Diaz-Santos2017}. Since CO observations for this sample are
not available, we estimated $M_{\mathrm{mol}}$, given the sSFR of each
source, considering the relation between the depletion time and sSFR
of galaxies. In the left panel we report the estimates obtained by
averaging the trend reported by \citet{Sargent2014} and
\citet{Scoville2017}, whereas in the right panel we report the
estimates obtained considering the trend by \citet{Scoville2017} only
(see Section \ref{subsec:cii_ssfr} for a more detailed discussion).}
\label{fig:acii}
\end{figure*}

In the local Universe, the majority of main-sequence galaxies have \cii~ luminosities that scale linearly with their IR luminosity
showing a constant $L_{\mathrm{\cii}}/L_{\mathrm{IR}}$ ratio, although
substantial scatter is present (e.g., \citealt{Stacey1991},
\citealt{Malhotra2001}, \citealt{Stacey2010}, \citealt{Cormier2015}, \citealt{Smith2017}). However, local (U)LIRGs appear to have a
different behaviour: they are typically \cii~ deficient with respect to their
IR luminosity, namely they have lower
$L_{\mathrm{\cii}}/L_{\mathrm{IR}}$ ratios than main-sequence galaxies (e.g.,
\citealt{Malhotra1997}, \citealt{Diaz-Santos2013},
\citealt{Farrah2013}). Furthermore, the $L_{\mathrm{\cii}}/L_{\mathrm{IR}}$ ratio correlates with the dust temperature, with the ratio decreasing for more luminous galaxies that have higher dust temperature (e.g. \citealt{Malhotra2001}, \citealt{Diaz-Santos2013}, \citealt{Gullberg2015}, \citealt{Diaz-Santos2017}). This relation also implies that $L_{\mathrm{\cii}}/L_{\mathrm{IR}}$ correlates  with $\langle U\rangle$, as the dust temperature is  proportional to the intensity of the radiation field ($\langle U\rangle \propto T_{\rm dust}^{4+\beta}$; e.g., Magdis et al. 2012). 
It is now well established that for main-sequence galaxies the dust temperature is rising with redshift (\citealt{Magdis2012}, \citealt{Bethermin2015}, \citealt{Schreiber2017}, following the trend $(1+z)^{1.8}$), as well as their IR luminosity, and sSFR. Our sample is made of $z \sim 2$ main-sequence galaxies, with SFRs comparable to those of (U)LIRGs and average $\langle U\rangle$ seven times larger that that of local spirals with comparable mass. Therefore, if the local relation between the $L_{\mathrm{\cii}}/L_{\mathrm{IR}}$ ratio and the dust temperature (and/or the IR luminosity, and/or the sSFR) holds even at higher redshift, we would expect our sample to be \cii~ deficient, showing a \cii -to-IR luminosity ratio similar to that of local (U)LIRGs.
 
To investigate this, we compare the \cii~ and IR luminosity of our sources with a compilation of measurements from the
literature in Figure \ref{fig:lcii_lir}. Our sample shows a
$L_{\mathrm{\cii}}/L_{\mathrm{IR}}$ ratio comparable to that observed
for local main-sequence sources (\citealt{Brauher2008},
\citealt{Cormier2015}, \citealt{Accurso2017}, \citealt{Contursi2017}), although it is shifted toward higher IR
luminosities as expected, given the higher SFR with respect to local
galaxies. The average $L_{\mathrm{\cii}}/L_{\mathrm{IR}}$ ratio of our data is $\sim 1.9 \times 10^{-3}$,
and has a scatter of $\sim$ 0.15 dex, consistent with the subsample of
$z \sim 1$ -- 2 main-sequence galaxies from \citet[filled grey stars
in Figure \ref{fig:lcii_lir}]{Stacey2010}. The $z \sim$ 1.8 sample of
\cite{Brisbin2015} is showing even higher ratios, surprisingly larger than all the
other literature samples at any redshift and IR luminosity. The \cii~
fluxes of these galaxies were obtained from ZEUS data and
ALMA observations will be needed to confirm them.
At fixed $L_{\mathrm{IR}}$ our galaxies show higher
$L_{\mathrm{\cii}}/L_{\mathrm{IR}}$ ratios than the average of the local IR-selected starbursts by \citet[2017]{Diaz-Santos2013}. The $L_{\mathrm{\cii}}/L_{\mathrm{IR}}$ ratio of our sample is also higher than that of the intermediate redshift starbursts from \cite{Magdis2014} and the subsample of $z \sim 1$ -- 2 starbursts from \citet[empty grey stars in Figure \ref{fig:lcii_lir}]{Stacey2010}.  This suggests that main-sequence galaxies have similar $L_{\mathrm{\cii}}/L_{\mathrm{IR}}$ ratios independently of their redshift and stellar mass, and points toward the conclusion that the $L_{\mathrm{\cii}}/L_{\mathrm{IR}}$ ratio is mainly set by the mode of star-formation (major mergers for starbursts and smooth accretion in extended disks for main-sequence galaxies), as suggested by \cite{Stacey2010} and \cite{Brisbin2015}.

We already knew that $L_{\mathrm{\cii}}$ does not universally scale with $L_{\mathrm{IR}}$, simply because of the existence of the \cii\ deficit. 
However, our results now also imply that the
$L_{\mathrm{\cii}}/L_{\mathrm{IR}}$ ratio does not only depend on
$L_{\mathrm{IR}}$: our $z=2$ main-sequence galaxies have similar
$L_{\mathrm{IR}}$ as local (U)LIRGs, but they have brighter \cii. For
similar reasons we can then conclude that the
$L_{\mathrm{\cii}}/L_{\mathrm{IR}}$ ratio does not depend on the dust
temperature, sSFR, or intensity of the radiation field only, and if
such relations exist they are not fundamental, as they depend at least
on redshift and likely on galaxies' star formation mode
(e.g. merger-driven for starbursts, or maintained by secular processes
for main-sequence galaxies). In Figure \ref{fig:u} we show the
relation between the $L_{\mathrm{\cii}}/L_{\mathrm{IR}}$ ratio and the
intensity of the radiation field for our sample and other local and
high-redshift galaxies from the literature. 

We note that $\langle U\rangle$ has been estimated in different ways
for the various samples reported in Figure \ref{fig:u}, depending on
the available data and measurements, and therefore some systematics
might be present when comparing the various datasets. In particular,
for our galaxies and those from \citet{Cormier2015} and
\citet{Cormier2017} it was obtained through the fit of the IR SED, as
detailed in Section
\ref{subsec:continuum_measurements} and \cite{Remy-Ruyer2014}. \citet{Diaz-Santos2017} and
\citet{Gullberg2015} instead do not provide an estimate of $\langle
U\rangle$, but only report the sources' flux at 63 $\mu$m and 158
$\mu$m \citep[R$_{64-158}$]{Diaz-Santos2017} and the dust temperature
\citep[T$_{\mathrm{dust}}$]{Gullberg2015}. Therefore we generated
\cite{Draine2007} models with various $\langle U\rangle$ in the range
2 -- 200 and fitted them with a modified black body template with
fixed $\beta$ = 2.0 (the same as used in the SED fitting for our
sample galaxies). We used them to find the following relations between
$\langle U\rangle$ and R$_{64-158}$ or T$_{\mathrm{dust}}$ and to
estimate the radiation field intensity for these datasets: $\log <U>=
1.144+1.807 \log R_{64-158}+0.540(\log R_{64-158})^2$ and $\log <U> =
−10.151 + 7.498 \log T_{\mathrm{dust}}$. Finally for the galaxies by
\citet{Capak2015} we used the relation between $\langle U\rangle$ and
redshift reported by \citet{Bethermin2015}.

The local galaxies of \citet{Diaz-Santos2017} indeed show a decreasing \cii -to-IR luminosity ratio with increasing $\langle U\rangle$ and the linear fit of this sample yields the following relation
\begin{equation}
\log (L_{\mathrm{\cii}}/L_{\mathrm{IR}}) = -2.1(\pm 0.1) + 0.7(\pm 0.1) \log(<{\mathrm{U}}>)
\end{equation}
and a dispersion of 0.3 dex.
However, high-redshift sources and local dwarfs deviate from the above relation, indicating that the correlation between $L_{\mathrm{\cii}}/L_{\mathrm{IR}}$ and $\langle U\rangle$ is not universal, but it also depends on other physical quantities, like redshift and/or galaxies' star formation mode. Our high-redshift main-sequence galaxies in fact show similar radiation field intensities as local (U)LIRGs, but typically higher $L_{\mathrm{\cii}}/L_{\mathrm{IR}}$ ratios. This could be due to the fact that in the formers the star formation is spread out in extended disks driving to less intense star-formation and higher $L_{\mathrm{\cii}}/L_{\mathrm{IR}}$, whereas in in the latters the star-formation, collision-induced by major mergers, is concentrated in smaller regions, driving to more intense star formation and lower $L_{\mathrm{\cii}}/L_{\mathrm{IR}}$, as suggested by \cite{Brisbin2015}.

This also implies that, since $L_{\mathrm{\cii}}/L_{\mathrm{IR}}$ does
not only depend on the intensity of the radiation field, and $\langle
U\rangle \propto M_{\mathrm{dust}}/L_{\mathrm{IR}}$, then
$L_{\mathrm{\cii}}$ does not simply scale with $M_{\mathrm{dust}}$
either\footnote{We note that the intensity of the radiation
    field $\langle U\rangle$ that we use for our analysis is different
    from the incident far-UV radiation field (G$_0$) that other
    authors report (e.g. \citealt{Abel2009}, \citealt{Stacey2010},
    \citealt{Brisbin2015}, \citealt{Gullberg2015}). However, according to PDR modelling, increasing the number of ionizing photons (G$_0$), more hydrogen atoms are ionized and the gas opacity decreases (e.g. \citealt{Abel2009}). More photons can therefore be absorbed by dust, and the dust temperature increases. As the radiation field's intensity depends on the dust temperature ($\langle U\rangle \propto T_{\mathrm{dust}}^{\alpha}$), then $\langle U\rangle$ is expected to increase with G$_0$ as well.}.

\subsection{\cii~ as a tracer of molecular gas}
\label{subsec:aCII}

Analogously to what discussed so far, by using a sample of local sources and distant starburst galaxies \cite{Gracia-Carpio2011} 
showed that starbursts show a similar \cii~ deficit at any time, 
but at high redshift the knee of the
$L_{\mathrm{\cii}}/L_{\mathrm{IR}}$ -- $L_{\mathrm{IR}}$ relation is shifted toward higher
IR luminosities, and a universal relation including all local and distant galaxies could be obtained by 
plotting the \cii\ (or other lines) deficit versus the star formation efficiency (or analogously their depletion time  $t_{\mathrm{dep}}$ = 1/SFE).

With our sample of $z=2$ main-sequence galaxies in hand, 
we would like now to proceed a step forward, and test whether 
 the \cii~ luminosity might be used as a tracer of
molecular gas mass: $L_{\mathrm{\cii}}\propto M_{\mathrm{mol}}$. In this case the $L_{\mathrm{\cii}}/L_{\mathrm{IR}}$ ratio would just be proportional to $M_{\mathrm{mol}}/SFR$ (given that $L_{\mathrm{IR}} \propto SFR$) and thus it would measure the galaxies' depletion time. The \cii~ deficit in starburst and/or mergers would therefore just reflect their shorter depletion time (and enhanced SFE) with respect to main-sequence galaxies.

In fact, the average $L_{\mathrm{\cii}}/L_{\mathrm{IR}}$ ratio of our $z \sim 2$ galaxies is $\sim 1.5$ times lower than the average of local main-sequence sources, consistent with the modest decrease of the depletion time from $z \sim 0$ to $z \sim 2$ (\citealt{Sargent2014}, \citealt{Genzel2015}, \citealt{Scoville2017}).
Although the scatter of the local and high-redshift measurements of the \cii~ and IR luminosities make this estimate quite noisy, this seems to indicate once more that the \cii~ luminosity correlates with the galaxies' molecular gas mass.

To test if this is indeed the case, as a first step we
complemented our sample with all literature data we could assemble (both main-sequence and starburst sources at low and high
redshift) with available \cii~ and molecular gas
mass estimates from other commonly used tracers (see the Appendix for details). 

We find that indeed L$_{\mathrm{\cii}}$ and
M$_{\mathrm{mol}}$ are linearly correlated, indepently of their main-sequence or starburst nature, and follow the relation
\begin{equation}
\log L_{\mathrm{\cii}} = -1.28(\pm 0.21) + 0.98(\pm 0.02) \log M_{\mathrm{mol}}
\end{equation}
with a dispersion of 0.3 dex (Figure \ref{fig:acii}). The Pearson test yields a 
coefficient $\rho = 0.97$, suggesting a statistically significant
correlation between these two parameters.

Given the linear correlation between the \cii~ luminosity and the
molecular gas mass, we can constrain
the $L_{\mathrm{\cii}}$-to-H$_2$ conversion factor.
In the following we refer to it as 
\begin{equation}
\alpha_{\mathrm{\cii}} = L_{\mathrm{\cii}}/M_{\mathrm{mol}}
\end{equation} 
by analogy
with the widely used CO-to-H$_2$ conversion factor,
$\alpha_{\mathrm{CO}}$. In Figure \ref{fig:acii} we report
$\alpha_{\mathrm{\cii}}$ as a function of redshift. 
\bigskip
\\
\textbf{Main-sequence galaxies}
\\
Considering only
the data available for main-sequence galaxies, we
get a median $\alpha_{\mathrm{\cii}} = 31$
M$_\odot$/L$_\odot$ with a median absolute deviation of 0.2 dex (and a standard deviation of 0.3 dex). We also computed the median
$\alpha_{\mathrm{\cii}}$ separately for the low- and high- redshift main-sequence
samples (Table \ref{tab:acii_median}): the two consistent estimates that we
obtained suggest that the \cii -to-H$_2$ conversion factor is likely
invariant with redshift. Furthermore, the medians of individual galaxies samples (green symbols in Figure \ref{fig:acii}) differ less than a factor 2 from one another and are all consistent with the estimated values of $\alpha_{\mathrm{\cii}} \sim 30$ M$_\odot$/L$_\odot$. 
\bigskip
\\
\textbf{Starburst galaxies}
\\
To further test the possibility to use the estimated $\alpha_{\mathrm{\cii}}$ not only for
main-sequence sources, but also for starbursts we 
considered the sample observed with the South Pole
Telescope (SPT) by \cite{Vieira2010} and \cite{Carlstrom2011}. They
are strongly lensed, dusty, star-forming galaxies at redshift $z \sim$ 2 -- 6 selected on the
basis of their bright flux at mm wavelengths (see Section
\ref{subsec:literature} for more details). \cii~
\citep{Gullberg2015} and CO \citep{Aravena2016}
observations are available for these targets. As \cite{Gullberg2015} notice, the similar \cii~ and CO line
velocity profiles suggest that these emission lines are likely not
affected by differential lensing and therefore their fluxes can be
directly compared. 
We obtained a median
$\alpha_{\mathrm{[CII]}} = 22$ M$_\odot$/L$_\odot$ for this sample, consistent
with that obtained for main-sequence datasets at both low and high redshift, as shown in Figure \ref{fig:acii}. As this
SPT sample is likely a mix of main-sequence and starburst galaxies \citep{Weiss2013}, we suggest that the
\cii -to-H$_2$ conversion factor is unique and independent of the source mode of star formation.

Similarly, we considered the starbursts at $z \sim 0.2$ analyzed by
\cite{Magdis2014} with available \cii~ and CO observations and the
sample of main-sequence and starbursts from the VALES survey \cite{Hughes2017}. 
The $M_{\mathrm{mol}}/L_{\mathrm{\cii}}$ ratios of these samples are
on average consistent with that of local and high-redshift
main-sequence galaxies, as shown in Figure \ref{fig:acii}.

Finally, we complemented our sample with the local galaxies observed
by \cite{Diaz-Santos2017} that are, in great majority,
(U)LIRGs. Molecular gas masses have not been published for these
sources and CO observations are not available. Therefore we estimated
$M_{\rm mol}$ considering the dependence of galaxies' depletion time
on their specific star formation rate, as parametrized by
\cite{Sargent2014} and \cite{Scoville2017}. Given the difference of
the two models especially in the starburst regime (see Section
\ref{subsec:cii_ssfr}), we estimated the gas masses for this sample
(i) adopting the mean depletion time obtained averaging the two
models, and (ii) considering the model reported by \cite{Scoville2017}
only. We report the results in Figure \ref{fig:acii} and
\ref{fig:cii_ssfr} (left and right bottom panels). If we adopt the gas
masses obtained with the first method, the $\alpha_{\rm \cii}$
conversion factor decreases by 0.3 dex for the most extreme
starbursts, whereas if only the model by \cite{Scoville2017} is
considered the $\alpha_{\rm \cii}$ conversion factor remains constant
independently of the main-sequence or starburst behaviour of galaxies
(see also Figure \ref{fig:cii_ssfr}, bottom panels). More future observations will be needed to explore in a more robust way the most extreme starburst regime.

All in all our results support the idea that the $\alpha_{\mathrm{\cii}}$ conversion factor is the same for main-sequence sources and starbursts, although the gas conditions in these two galaxy populations are different (e.g. starbursts have higher gas densities and harder radiation fields than main-sequence galaxies). Possible reasons why, despite the different conditions, \cii~ correlates with the molecular gas mass for both populations might include the following: (i) different parameters might impact the $L_{\mathrm{\cii}}/M_{\mathrm{mol}}$ ratio in opposite ways and balance, therefore having an overall negligible effect; (ii) the gas conditions in the PDRs
might be largely similar in all galaxies, with variations in the \cii/CO ratio smaller than a factor $\sim 2$ and most of the \cii~ produced in the molecular ISM (\citealt{deLooze2014}, \citealt{Hughes2015}, \citealt{Schirm2017}).

\vspace{0.5 cm}
Finally, we investigated what is the main reason for the scatter of the $\alpha_{\mathrm{\cii}}$ measurements. We considered only the galaxies with $M_{\rm mol}$ determined homogeneously from the CO luminosity and we estimated the scatter of the $L_{\rm \cii}$ - $L'_{\rm CO}$ relation.  The mean absolute deviation of the relation is $\sim$ 0.2 dex, similar to that of the $L_{\rm \cii}$ - $M_{\rm mol}$ relation. This is mainly due to the fact that, to convert the CO luminosity into molecular gas mass, commonly it is adopted an $\alpha_{\rm CO}$ conversion factor that is very similar for all galaxies (it mainly depends on metallicity and the latter is actually very similar for all the galaxies that we considered as shown in Figure \ref{fig:cii_metall}). More interestingly, the mean absolute deviation of the $L_{\rm \cii}$ - $L'_{\rm CO}$ relation is comparable to that of $\alpha_{\rm \cii}$. We therefore concluded that the scatter of the \cii -to-molecular gas conversion factor is mainly dominated by the intrinsic scatter of the \cii -to-CO luminosity relation, although the latter correlation is not always linear (e.g. see Figure 2 in \citealt{Accurso2017}) likely due to the fact that \cii~ traces molecular gas even in regimes where CO does not.

\subsection{The dependence of the \cii -to-IR ratio on galaxies' distance from the main-sequence}
\label{subsec:cii_ssfr}

As the next step, we explicitly investigated if indeed $L_{\mathrm{\cii}}\propto M_{\mathrm{mol}}$, 
when systematically studying galaxies on and off main-sequence, thus spanning a large range of 
sSFR and SFE, up to merger-dominated systems.
In fact, when comparing low- and high-redshift sources in bins of IR
luminosity (Figure \ref{fig:lcii_lir}) we might be mixing, in each bin, galaxies with very
different properties (e.g. high-$z$ main-sequence sources with local starbursts). On the contrary, this does not happen when considering bins of distance from the main-sequence (namely, sSFR/sSFR$_{\mathrm{MS}}$).

\begin{figure*}
\centering
\includegraphics[width=\textwidth]{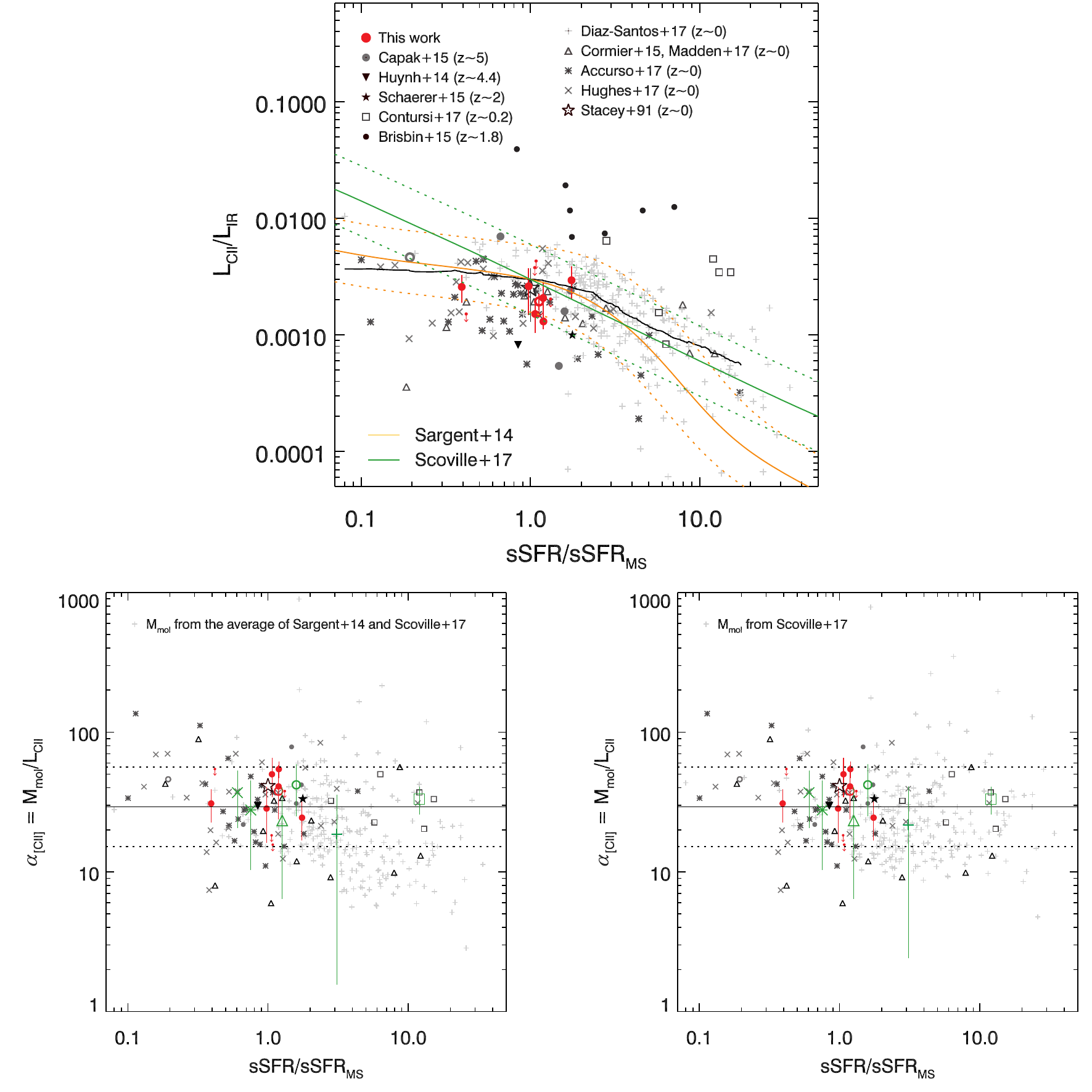}
\caption{Correlation between the \cii~ luminosity and galaxies' distance from the main-sequence. Top panel: \cii -to-IR luminosity ratio as
  a function of the galaxy distance from the main sequence. The
  symbols are the same as reported in Figure \ref{fig:lcii_lir}
  caption. Additionally, we include the average of the
  local star-forming galaxies from \citet[cyan star]{Stacey1991}. In
  particular, the sources by \citet{Brisbin2015} might be lensed, but the
  magnification factors are unknown and therefore we plot the observed
  values. We
  also show the running mean computed considering all the plotted datapoints a part from the sample from \citet[black solid line]{Contursi2017}. Finally we report the
  model by \citet[yellow curve]{Sargent2014} and \citet[green
  curve]{Scoville2017}, showing the trend of the depletion time as a
  function of the sSFR, renormalized to match the observed
  $L_{\mathrm{\cii}}/L_{\mathrm{IR}}$ ratios (the standard deviations
  of the models are marked as dashed curves). Bottom panels: dependence of $\alpha_{\mathrm{\cii}}$ from galaxies' distance from the main-sequence. The difference between the left and
right panels concerns how the molecular gas mass was estimated for the
sample of local galaxies from \citet[light gray
crosses]{Diaz-Santos2017}. Since CO observations for this sample are
not available, we estimated $M_{\mathrm{mol}}$, given the sSFR of each
source, considering the relation between the depletion time and sSFR
of galaxies. In the left panel we report the estimates obtained by
averaging the trend reported by \citet{Sargent2014} and
\citet{Scoville2017}, whereas in the right panel we report the
estimates obtained considering the trend by \citet{Scoville2017} only
(see Section \ref{subsec:cii_ssfr} for a more detailed discussion).}
\label{fig:cii_ssfr}
\end{figure*}

We considered  samples with available sSFR measurements and in Figure \ref{fig:cii_ssfr} we plot the
$L_{\mathrm{\cii}}/L_{\mathrm{IR}}$ ratio in bins of
sSFR, normalized to the sSFR of the main-sequence at each 
redshift \citep{Rodighiero2014}. Our sample has a
$L_{\mathrm{\cii}}/L_{\mathrm{IR}}$ ratio comparable to that reported in the
literature for main-sequence galaxies at lower (\citealt{Stacey1991},
\citealt{Cormier2015}, the subsample of main-sequence galaxies from \citealt{Diaz-Santos2017}) and higher redshift
(\citealt{Capak2015})\footnote{For this sample we derived $L_{\mathrm{IR}}$ from ALMA
  continuum using the main-sequence templates of
  \cite{Magdis2012} and an appropriate temperature 
  for $z=5.5$, following the evolution given in \citet{Bethermin2015} and \citet{Schreiber2017}. This is the reason why the values that we are plotting differ from those published by \cite{Capak2015}, but are equivalent to those recently revised by \cite{Brisbin2017}.}. This is
up to $\sim 10$ times higher
than the typical $L_{\mathrm{\cii}}/L_{\mathrm{IR}}$ ratio of
starbursts defined as to fall $> 4$ times above the main-sequence
\citep{Rodighiero2011}. Given the fact that the IR luminosity is
commonly used as a SFR tracer and the \cii~ luminosity seems to
correlate with the galaxies' molecular gas mass, we expect the
$L_{\mathrm{\cii}}/L_{\mathrm{IR}}$ ratio to depend on galaxies' gas depletion
time ($\tau _{\mathrm{dep}} = M_{\mathrm{mol}}/SFR$). This seems to be
substantiated by the fact that
the depletion time in main-sequence galaxies is on average $\sim$ 10
times higher than in starbursts (e.g. \citealt{Sargent2014},
\citealt{Scoville2017}), similarly to what is observed for the
$L_{\mathrm{\cii}}/L_{\mathrm{IR}}$ ratio. To make this comparison
more quantitative, we
considered two models (\citealt{Sargent2014}, \citealt{Scoville2017}) predicting how
the depletion time of galaxies changes as a function of their distance
from the main-sequence and rescaled them to match the 
$L_{\mathrm{\cii}}/L_{\mathrm{IR}}$ observed for main-sequence galaxies. This scaling factor mainly depends on the \cii -to-CO luminosity ratio and given the shift we applied to the \cite{Sargent2014} and \cite{Scoville2017} models we estimated $L_{\mathrm{\cii}}/L_{\mathrm{CO}} \sim 6000$. This is in good agreement with the typical values reported in the literature and ranging between 2000 -- 10000 (\citealt{Stacey1991}, \citealt{Magdis2014}, \citealt{Accurso2017}, \citealt{Rigopoulou2017}). We compare the rescaled models with observations in Figure \ref{fig:cii_ssfr}. Given the higher number of main-sequence sources than starbursts, uncertainties on the estimate of stellar masses
affecting galaxies sSFR would tend to systematically bias the
distribution of $L_{\mathrm{\cii}}/L_{\mathrm{IR}}$ towards
higher ratios as the distance from the main-sequence increases (similarly
to the Eddington bias affecting source luminosities in surveys). To take
this observational bias into account, we convolved the models by
\cite{Sargent2014} and \cite{Scoville2017} with a Gaussian function with FWHM $\sim$ 0.2 dex
(the typical uncertainty affecting stellar masses). Qualitatively, the drop of the depletion time that both models show with increasing sSFR well reproduces the trend of the \cii -to-IR luminosity ratio with sSFR/sSFR$_{\rm MS}$ that is observed in Figure \ref{fig:cii_ssfr}. Considering that $\tau _{\mathrm{dep}} = M_{\mathrm{mol}}$/SFR, and that the
IR luminosity is a proxy for the SFR, the agreement between models and
observations suggests that \cii~ correlates reasonably well with the molecular gas
mass, keeping into account the limitations of this exercise (there are still lively ongoing debates on how to best estimate the gas mass of off main-sequence galaxies, as reflected in the differences in the models we adopted). In this framework, the \cii~ deficiency of starbursts can be
explained as mainly due to their higher
star formation efficiency, and hence far-UV fields, with respect to main-sequence sources. 
This is consistent with the invariance found by \cite{Gracia-Carpio2011}, but it conceptually extends it to the possibility that \cii\ is directly proportional to the molecular gas mass, at least empirically. 

\begin{figure*}
\includegraphics[width=\textwidth]{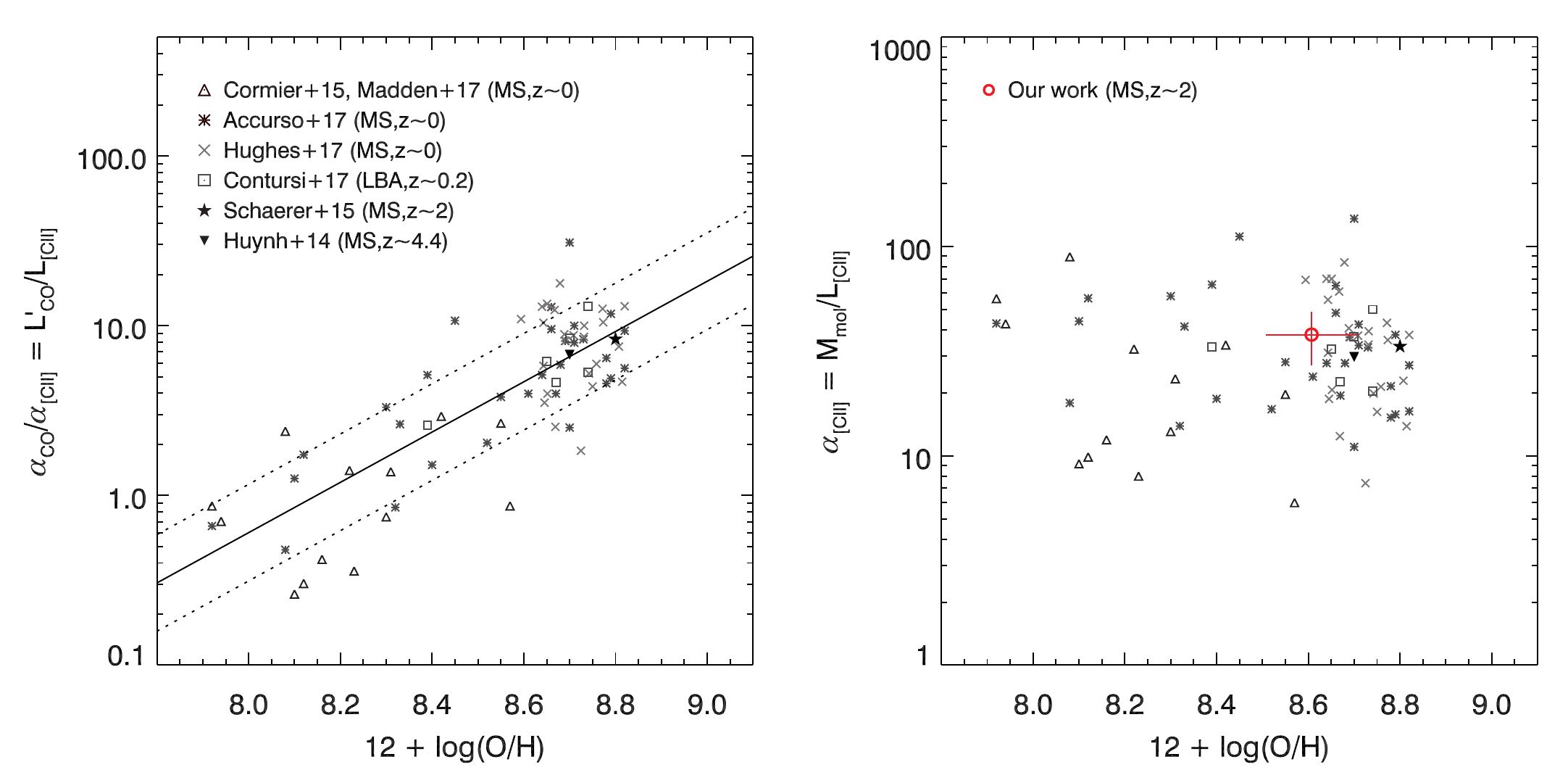}
\caption{Metallicity dependence of $\alpha_{\mathrm{\cii}}$ for multiple samples with available metallicity estimate, all  homogenized to the \citet{Pettini2004} calibration using the parametrizations by \citet{Kewley2008}. The symbols are the same as reported in Figure \ref{fig:lcii_lir} caption  and legend. Left panel: ratio of the CO and \cii~ luminosity as a function of the galaxies gas phase metallicity. The linear fit of the two samples is reported (black solid line). Right panel: \cii -to-H$_2$ conversion factor as a function of metallicity. The average measurement for our sample (red empty circle) is reported only in this panel since no CO measurements are available for our sources. The gas mass for our galaxies was estimated considering the integrated Schmidt-Kennicutt relation (see Section \ref{subsec:continuum_measurements}).  We note that one of the galaxies by \citet{Cormier2015} is an outlier to the $L_{\rm \cii}$ -- $M_{\rm mol}$ relation (and therefore of the $\alpha_{\rm \cii}$ -- metallicity estimate) due to its very low \cii~ luminosity with respect to the CO one. We kept this galaxy in the sample for consistency with the literature, although there might be some issues with its \cii~ and/or CO measurements.}
\label{fig:cii_metall}
\end{figure*}

However, quantitatively some discrepancies between models and observations are present. The model by \cite{Sargent2014} accurately reproduces observations, at least up to sSFR/sSFR$_{\mathrm{MS}} \sim 4$, but some inconsistencies are found at high sSFR/sSFR$_{\mathrm{MS}}$. On the contrary, the model by \cite{Scoville2017} reproduces the observations for galaxies on and above the main-sequence, even if some discrepancies are present at sSFR/sSFR$_{\mathrm{MS}} < 1$, a regime that is not yet well tested (but see \citealt{Schreiber2017b}, \citealt{Gobat2017b}). 
 Some possible explanations for the discrepancy between the observations and the model by \cite{Sargent2014} are the following: (i) starbursts might have higher gas
fractions than currently predicted by the \cite{Sargent2014} model, in agreement with the \cite{Scoville2017} estimate; (ii) the \cii~ luminosity, at
fixed stellar mass, is expected to increase with more intense radiation
fields such as those characteristics of starbursts
(\citealt{Narayanan2017}, \citealt{Diaz-Santos2017}, \citealt{Cormier2017}), possibly leading to too high \cii -to-IR luminosity ratios with respect to the model by \cite{Sargent2014}; (iii) if the fraction of \cii~ emitted by molecular gas decreases when the sSFR increases (e.g. for starbursts) as indicated by the model from \cite{Accurso2017b}, then the \cii -to-IR luminosity ratio would be higher than the expectations from the model by \cite{Sargent2014}. However, to reconcile the observations of the most extreme starbursts (sSFR/sSFR$_{\mathrm{MS}} \sim 10$) with the model, the \cii~ fraction emitted from molecular gas should drop to $\sim$ 30\%, which is much lower than the predictions from \cite{Accurso2017b}; (iv) we might also be facing an
observational bias: starbursts with
relatively high \cii~ luminosities might have been preferentially observed
so far. Future deeper observations will allow us to understand if this
mismatch is indeed due to an observational bias or if instead 
is real. In the latter case it would show that
$\alpha_{\mathrm{\cii}}$ is not actually constant in the strong
starburst regime. 

We also notice that some local Lyman break analogs observed by \cite{Contursi2017} show $L_{\mathrm{\cii}}/L_{\mathrm{IR}}$ ratios higher than expected from both models, given their sSFR (Figure \ref{fig:cii_ssfr}). Although these sources have sSFRs typical of local starbursts, their SFEs are main sequence-like as highlighted by \cite{Contursi2017}. They are likely exceptional sources that do not follow the usual relation between sSFR and SFE. Given the fact that they show \cii -to-IR luminosity ratios compatible with the average of main-sequence galaxies (Figure \ref{fig:lcii_lir}), we conclude that also in this case the SFE is the main parameter setting $L_{\mathrm{\cii}}/L_{\mathrm{IR}}$, suggesting that the \cii~ luminosity correlates with galaxies' molecular gas mass. 

\subsection{Invariance of $\alpha_{\mathrm{\cii}}$ with gas phase metallicity}
\label{subsec:metall}

In this Section we investigate the dependence of the $\alpha_{\mathrm{\cii}}$ conversion factor on gas phase metallicity. Understanding whether \cii\ traces the molecular gas also for low metallicity galaxies is relevant for observations of high-redshift galaxies that are expected to be metal-poor (\citealt{Ouchi2013}; \citealt{Vallini2015}). 

In Figure \ref{fig:cii_metall} we show literature
samples with available measurements of metallicity, CO, and \cii~
luminosities. To properly compare different samples we converted all metallicity estimates to the calibration by \cite{Pettini2004} using the parametrizations by \cite{Kewley2008}.
We converted the CO luminosity into gas mass by assuming the following $\alpha_{\mathrm{CO}}$ -- metallicity dependence:
\begin{equation}
\log \alpha_{\mathrm{CO}} = 0.64 - 1.5 (12 + \log (O/H) - 8.7) 
\end{equation}
that yields the Galactic $\alpha_{\mathrm{CO}}$ for solar metallicities and has a slope in between those found in the literature (typically ranging between -1 and -2, e.g. \citealt{Genzel2012}, \citealt{Schruba2012}, \citealt{Tan2014}, \citealt{Accurso2017}, \citealt{Sargent2017}). Adopting an $\alpha_{\mathrm{CO}}$ -- metallicity dependence with a slope of $-1$ or $-2$ instead would not change our conclusions.

We show the ratio between the CO and \cii~ luminosity as a function of metallicity in Figure \ref{fig:cii_metall} (left panel). This plot was first shown by \cite{Accurso2017} (see their Figure 2)
and here we are adding some more literature datapoints. 
Over the metallicity range spanned by these samples
($12 + \log \mathrm{O/H} \sim 7.8$ -- 9), the CO luminosity drops by a factor 20 compared to \cii.  The fact that the $L'_{\rm CO}/M_{\rm dust}$ ratio is overall constant with metallicity (given that both the gas-to-dust ratio and $\alpha_{\rm CO}$ similarly depend on metallicity) implies that $L_{\cii}/M_{\rm dust}$ has large variations with metallicity (similarly to the $L_{\cii}/L_{\rm CO}$ ratio), consistent with what discussed in Section \ref{subsec:cii_deficit} (namely that \cii\ is not simply a dust mass tracer).

In Figure \ref{fig:cii_metall} (right panel) 
we show the $\alpha_{\mathrm{\cii}}$ dependence on metallicity. Although the scatter is quite large, the
$L_{\mathrm{\cii}}/M_{\mathrm{mol}}$ ratio does not
seem to depend on metallicity. When fitting the data with a linear function we obtain a slope of $-0.2 \pm 0.2$, which is not significantly different from zero and consistent with a constant relation, and a standard deviation of 0.3 dex. This suggests that \cii~ can be used as a
``universal'' molecular gas tracer and a particularly convenient
tool to empirically estimate the gas mass of starbursts (whose metallicity is notoriously
difficult to constrain due to their high dust extinction) and 
high-redshift low-metallicity galaxies.

We note that the \cii~ luminosity is expected to become fainter at very low metallicities, due to the simple fact that less carbon is present \citep{Cormier2015}. However, this effect is negligible for the samples that we are considering and likely only becomes important at very low metallicities (12 + log(O/H) $<$ 8.0).

\begin{table}
\center
\caption{Estimates of the \cii -to H$_2$ conversion factor.}
\label{tab:acii_median}
\begin{tabular}{c c c c c}
\toprule
\midrule
Samples & Mean & Standard deviation & Median & M.A.D.\\[3pt]
 & [M$_\odot$/L$_\odot$] & [dex] & [M$_\odot$/L$_\odot$] & [dex] \\[3pt]
(1) & (2) & (3) & (4) & (5)\\ [3pt]
\hline
\rule{0pt}{4ex}

All & 31 & 0.3 & 31 & 0.2 \\
Local & 30 & 0.3 & 28 & 0.2 \\
High-$z$ & 35 & 0.2 & 38 & 0.1 \\

\bottomrule
\end{tabular}
\begin{minipage}{9cm}
\smallskip
\footnotesize
\textbf{Columns} (1) Samples used to compute
$\alpha_{\mathrm{\cii}}$. For the local estimate we considered the
\cite{Accurso2017} and \cite{Cormier2015} datasets, whereas for the
high-redshift one we used our measurements together with those by
\cite{Capak2015}. The global estimate of
$\alpha_{\mathrm{\cii}}$ was done by considering all the
aforementioned samples.; (2) mean
$\alpha_{\mathrm{\cii}}$; (3) standard
deviation of the $\alpha_{\mathrm{\cii}}$ estimates; (4) median $\alpha_{\mathrm{\cii}}$; (5) mean absolute deviation of $\alpha_{\mathrm{\cii}}$ estimates.
\end{minipage}
\end{table}

\begin{figure*}
\includegraphics[width=\textwidth]{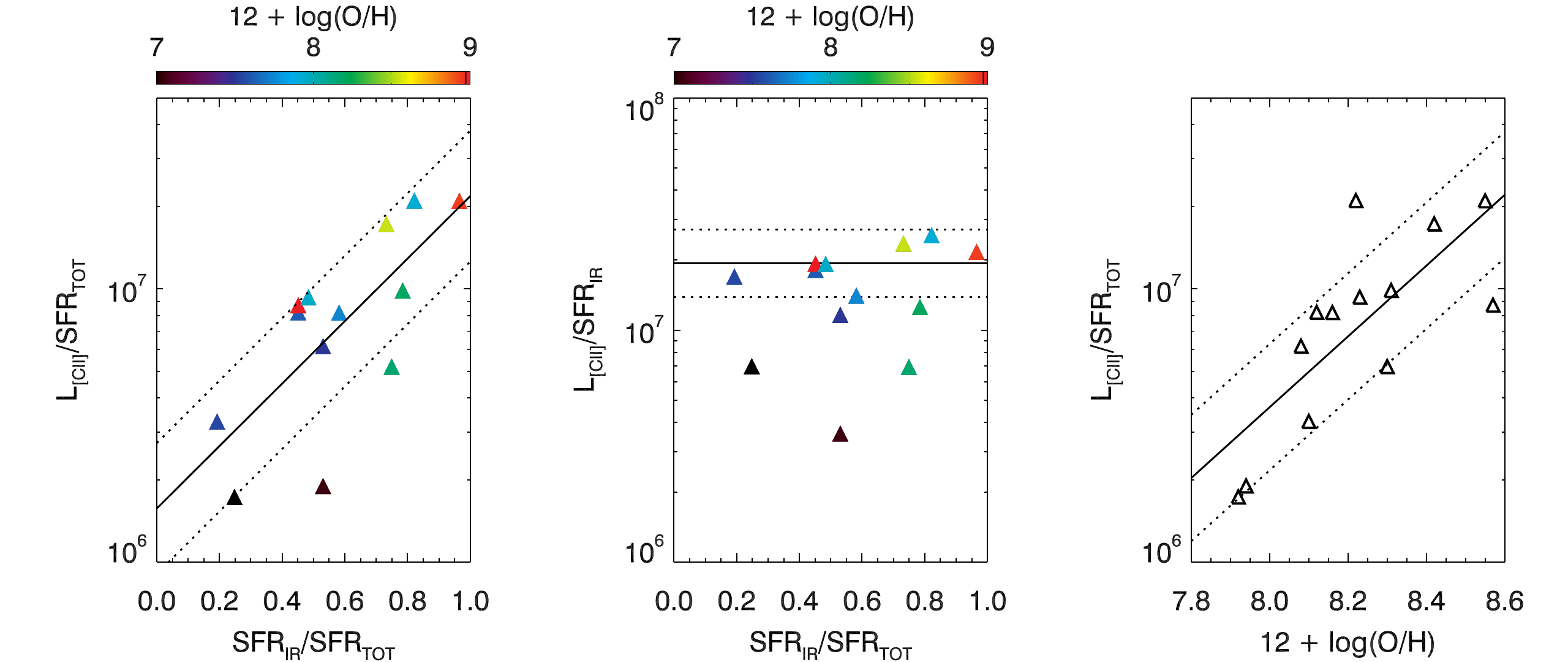}
\caption{\cii~ dependence on the galaxies' total (UV + IR) and
  obscured (IR only) SFR. The sample is made of the low-metallicity sources by \citet{Cormier2015}, \citet{Cormier2017}. Left panel: dependence of the \cii~ luminosity to total SFR ratio on the ratio between the total and obscured SFR. The fit of the data is reported (solid black line) together with the standard deviation of the data (dashed black line). Central panel: dependence of the \cii~ luminosity to obscured SFR ratio on the ratio between the total and obscured SFR. The average ratio for our $z\sim2$ sample of main-sequence galaxies is reported (solid black line) together with its uncertainty (dashed black line). Right panel: dependence of the \cii~ luminosity to total SFR on the gas phase metallicity. The fit of the data is reported (solid black line) together with the standard deviation of the data (dashed black line).}
\label{fig:sfr}
\end{figure*}

\subsection{Implications for surveys at $z > 2$}
\label{subsec:surveys}

As shown in the previous Sections, \cii~ correlates with the galaxies' molecular gas mass, and the \cii -to-H$_2$ conversion factor is likely
independent of the main-sequence and starburst behaviour of galaxies, as well as of
their gas phase metallicity. In perspective, this is
particularly useful for studies of high-redshift targets. At high
redshift in fact, due to the galaxies' low metallicity, CO is expected not to trace the bulk of the H$_2$ anymore (e.g. \citealt{Maloney1988}, \citealt{Madden1997}, \citealt{Wolfire2010}, \citealt{Bolatto2013}). Thanks to its high luminosity even in the low metallicity regime, \cii~ might become a
very useful tool to study the ISM properties at these
redshifts. However some caution is needed when interpreting or predicting the \cii~ luminosity at high redshift. 
Recent studies have shown that low-metallicity galaxies have low dust
content, hence the UV obscuration is minimal and the IR emission is
much lower than in high-metallicity sources
(e.g. \citealt{Galliano2005}, \citealt{Madden2006},
\citealt{Remy-Ruyer2013}, \citealt{deLooze2014},
\citealt{Cormier2015}). This means that the obscured star formation
rate -- that can be computed from the IR luminosity through the
calibration done by \cite{Kennicutt1998}  -- can be up to 10 times
lower than the unobscured one (e.g. computed thorough the UV SED
fitting). This can be seen also in Figure \ref{fig:sfr} where we report the sample of local low-metallicity galaxies from \cite{Cormier2015} and \cite{Cormier2017}, taking at face value the SFR estimates from the literature. The $SFR_{\rm IR}/SFR_{\rm TOT}$ ratio clearly depends on the galaxies' metallicity, with the most metal-poor showing on average lower ratios. Furthermore, the ratio between the \cii~ luminosity and the total SFR of these galaxies linearly depends on the $SFR_{\rm IR}/SFR_{\rm TOT}$ ratio (Figure \ref{fig:sfr}, left panel):
\begin{equation}
\log (L_{\rm \cii}/SFR_{\rm TOT}) = 6.2(\pm 0.2) +  1.1(\pm 0.3) SFR_{\rm IR}/SFR_{\rm TOT}
\end{equation}
with a scatter of 0.2 dex, indicating that galaxies with lower
metallicity (and lower obscured SFR) typically have lower $L_{\rm
  \cii}/SFR_{\rm TOT}$ ratios. This is clearly visible in Figure \ref{fig:sfr} (right panel): the dependence of the $L_{\rm  \cii}/SFR_{\rm TOT}$ ratio on metallicity can be parametrized as follows:
\begin{equation}
\log (L_{\rm  \cii}/SFR_{\rm TOT}) = -3.8(\pm 2.8) + 1.3(\pm 0.3) [12 + \log (O/H)]
\end{equation}
with a dispersion of 0.2 dex.
On the contrary, the ratio between the
\cii~ luminosity and the obscured SFR is constant with the $SFR_{\rm
  IR}/SFR_{\rm TOT}$ ratio (Figure \ref{fig:sfr}, central panel). This
suggests that the \cii~ emission is related to dusty star-forming
regions rather than to the whole SFR of the galaxy. At very high
redshift (e.g. $z>4$) measuring the IR luminosity is problematic and
therefore often the total SFR obtained from UV-corrected estimates is
used to derive a measurement of $L_{\rm IR}$. However, this might lead
to overestimate the IR luminosity and therefore bias the \cii -to-IR
luminosity ratio toward lower values. This would mean that the \cii~
deficit observed at high redshift might be due to the approximate
estimate of the IR luminosity and not only due to the real evolution of the ISM properties. It could also explain the several cases of $z > 5$ galaxies with \cii~ non-detections that have been recently reported (\citealt{Combes2012}, \citealt{Ouchi2013}, \citealt{Maiolino2015}, \citealt{Schaerer2015}, \citealt{Watson2015}): if the total SFR was used to estimate the $L_{\rm IR}$ and the typical $L_{\rm \cii}/L_{IR} = 2\times 10^{-3}$ ratio was used to predict the \cii~ luminosity when proposing for observing time, the $L_{\rm \cii}$ would have been overestimated and therefore the observations would have not been deep enough to detect the \cii~ emission of the targets. Future actual measurements of the IR luminosity will be crucial to assess whether high-redshift observations were biased, or on the contrary if the \cii~ deficiency is due to an actual evolution of galaxies' properties from $z\sim0$ to $z\sim5$. In the latter case the reason for the deficiency might still not be clear and an additional word of caution is needed: if the \cii~ luminosity traces the molecular gas mass even at these high redshifts, these sources might be \cii~ deficient due to a low molecular gas content and high SFE. However, the different conditions of the ISM at these redshifts, the lower dust masses, and likely the much harder radiation fields might play an important role as well, potentially introducing systematics and limiting the use of \cii~ as molecular gas tracer for very distant galaxies.

\subsection{Caveats}

Finally we mention a few caveats that it is important to
consider when using the \cii~ emission line to trace galaxies'
molecular gas.

First, as discussed in Section \ref{subsec:surveys}, at redshift
$z \gtrsim 5$ the ISM conditions are likely different with respect to
lower redshift (e.g. lower dust masses,
harder radiation fields). This might impact the \cii~
luminosity, possibly introducing some biases, and limiting the use of
the \cii~ emission line to estimate the molecular gas mass of galaxies
at very high redshift.

Secondly, there are local studies indicating that \cii, mainly due to its
low ionization potential, is simultaneously tracing the molecular, atomic and ionized phases (e.g. \citealt{Stacey1991},
\citealt{Sargsyan2012}, \citealt{Rigopoulou2014}, \citealt{Diaz-Santos2017}, \citealt{Croxall2017}). The total measured \cii~ luminosity might therefore be higher than the one arising from the molecular gas only: this would lead to overestimated H$_2$ masses. However, it seems that 70\% -- 95\%  of the \cii~ luminosity originates from PDRs (\citealt{Cormier2015}, \citealt{Diaz-Santos2017}) and in particular $> 75$\% arises from the molecular phase (\citealt{Pineda2013}, \citealt{Velusamy2014}, \citealt{Vallini2015}, \citealt{Olsen2017}, \citealt{Accurso2017b}). 

Lastly, as opposed to CO, \cii~ is likely
emitted only in regions where star formation is ongoing. Molecular clouds
that are not illuminated by young stars would therefore not be detected \citep{Beuther2014}. 

All in all, the limitations affecting \cii~ seem to be different
with respect to the ones having an impact on the molecular gas tracers
commonly used so far (CO, \ci, or dust
measurements), making it an independent molecular gas
proxy. Future works comparing the gas mass estimates obtained with
different methods will help understanding what tracer is better
to consider depending on the physical conditions of the target.

\section{Conclusions}
\label{sec:summary}

In this paper we discuss the analysis of a sample of 10 main-sequence
galaxies at redshift $z \sim 2$ in GOODS-S. We present new ALMA Band 7 850 $\mu$m observer frame continuum, and Band 9 \cii~ line together with 450 $\mu$m observer frame continuum observations, complemented by a suite of ancillary
data, including \textit{HST}, \textit{Spitzer}, \textit{Herschel}, and VLA imaging,
plus VLT and Keck longslit spectroscopy. The goal is to investigate
whether $z \sim 2$, main-sequence galaxies are \cii~ deficient and 
understand what are the main physical parameters affecting the \cii~ luminosity. We summarize in the following the main conclusions we reached.

\begin{itemize}[leftmargin=*]

\item The ratio between the \cii~ and IR luminosity
  ($L_{\mathrm{[CII]}}/L_{\mathrm{IR}}$) of $z \sim 2$  main-sequence galaxies is
  $\sim 2 \times 10^{-3}$, comparable to that of local main-sequence sources and a factor of
  $\sim 10$ higher than local starbursts. This implies that there is not a
  unique correlation between $L_{\mathrm{\cii}}$ and L$_\mathrm{IR}$
  and therefore we should be careful when using the \cii~ luminosity as a SFR indicator. Similarly, the \cii~ luminosity does not uniquely correlate with galaxies' specific star formation rate, intensity of the radiation field, and dust mass.

\item  The \cii\ emission is spatially extended, on average, on scales comparable to 
the stellar mass sizes (4 -- 7~kpc), as inferred from \textit{HST} imaging in the optical rest frame. This is in agreement with the results by \cite{Stacey2010}, \cite{Hailey-Dunsheath2010}, and \cite{Brisbin2015} who, for samples of $z \sim 1$ -- 2 galaxies, find similar \cii~ extensions. This also suggests that
our sample of main sequence galaxies, with typical stellar masses and
SFRs, is not made up of the ultra-compact (and more massive) sources selected and studied by \cite{Tadaki2015} and \cite{Barro2016}.

\item The \cii~ luminosity linearly correlates with galaxies' molecular gas masses. By complementing our sample with those from the literature, we constrained the $L_{\mathrm{\cii}}$-to-H$_2$ conversion factor: it has a median
  $\alpha_{\mathrm{\cii}} = 31$ M$_\odot$/L$_\odot$ and a median
  absolute deviation of $\sim$ 0.2 dex. We find it mostly invariant with galaxies' redshift, depletion time, and
  gas phase metallicity. This makes \cii~ a convenient emission
  line to estimate the gas mass of starbursts, a notoriously hard property to constrain by using the CO and dust emission due to the large uncertainties in the conversion factors to be adopted. Furthermore, the invariance of $\alpha_{\mathrm{\cii}}$ with metallicity together with the remarkable brightness of \cii~
  makes this emission line a useful tool to constrain
  gas masses at very high redshift, where galaxies' metallicity is expected
  to be low.

\item Considering that \cii~ traces the molecular gas and the IR luminosity is a proxy for SFR, the $L_{\rm \cii}/L_{\rm IR}$ ratio seems to be mainly a tracer of galaxies' gas depletion time. The $L_{\rm \cii}/L_{\rm IR}$ ratio for our sample of $z\sim2$ main-sequence galaxies is $\sim$ 1.5 times lower than that of local main-sequence samples, as expected from the evolution of depletion time with redshift.
 
 \item The weak \cii\ signal from $z>6$ -- 7 galaxies and the many non-detections in the recent literature might be evidence of high star formation efficiency, but might be also due to the fact that the expected signal is computed from the total UV star formation rate, while local dwarfs suggest that \cii\ only reflects the portion of SFR reprocessed by dust in the IR.

\item Although some caveats are present (e.g.
  \cii~ non-detections at very high redshift might also be due to the effects of a strong radiation field; \cii~ might be tracing
  different gas phases simultaneously; it is only emitted when the
  gas is illuminated by young stars, so it only traces molecular gas with ongoing star formation), the limitations that affect
  \cii~ are different with respect to those impacting more
  traditional gas tracers such as CO, \ci, and dust emission. This
  makes \cii~ an independent proxy, particularly
  suitable to push our current knowledge of galaxies' ISM to the
  highest redshifts.

\end{itemize}

\section*{Acknowledgments}
We are grateful to the anonymous referee for their insightful
comments.
A.Z. thanks C. Cicone, G. Accurso, A. Saintonge, Q. Tan, M. Aravena, A. Pope, A. Ferrara, S. Gallerani, and A. Pallottini for useful discussions. T.M.H. acknowledges support from the Chinese Academy of Sciences (CAS)
and the National Commission for Scientific and Technological Research
of Chile (CONICYT) through a CAS-CONICYT Joint Postdoctoral Fellowship
administered by the CAS South America Center for Astronomy (CASSACA)
in Santiago, Chile. D.C. is supported by the European Union's Horizon 2020 research
and innovation programme under the Marie Sk\l{}odowska-Curie
grant agreement No 702622. M.T.S was supported by a Royal Society Leverhulme Trust Senior Research Fellowship (LT150041). W.R. is supported by JSPS KAKENHI Grant Number JP15K17604 and the
Thailand Research Fund/Office of the Higher Education Commission Grant
Number MRG6080294. D. L. acknowledges funding from the European Research Council (ERC) under the European Union's Horizon 2020 research and innovation programme (grant agreement No. 694343).
This paper makes use of the
following ALMA data: ADS/JAO.ALMA\#2012.1.00775.S
ALMA is a partnership of European Southern Observatory
(ESO, representing its member states), NSF (USA)
and NINS (Japan), together with NRC (Canada), NSC and
ASIAA (Taiwan), and KASI (Republic of Korea), in cooperation
with the Republic of Chile. The Joint ALMA
Observatory is operated by ESO, AUI/NRAO and NAOJ.

\bibliographystyle{mnras} 
\bibliography{bibliography}

\appendix
\section{A. Resolution}
\label{app:resolution}

\begin{figure}
\includegraphics[width=0.5\textwidth]{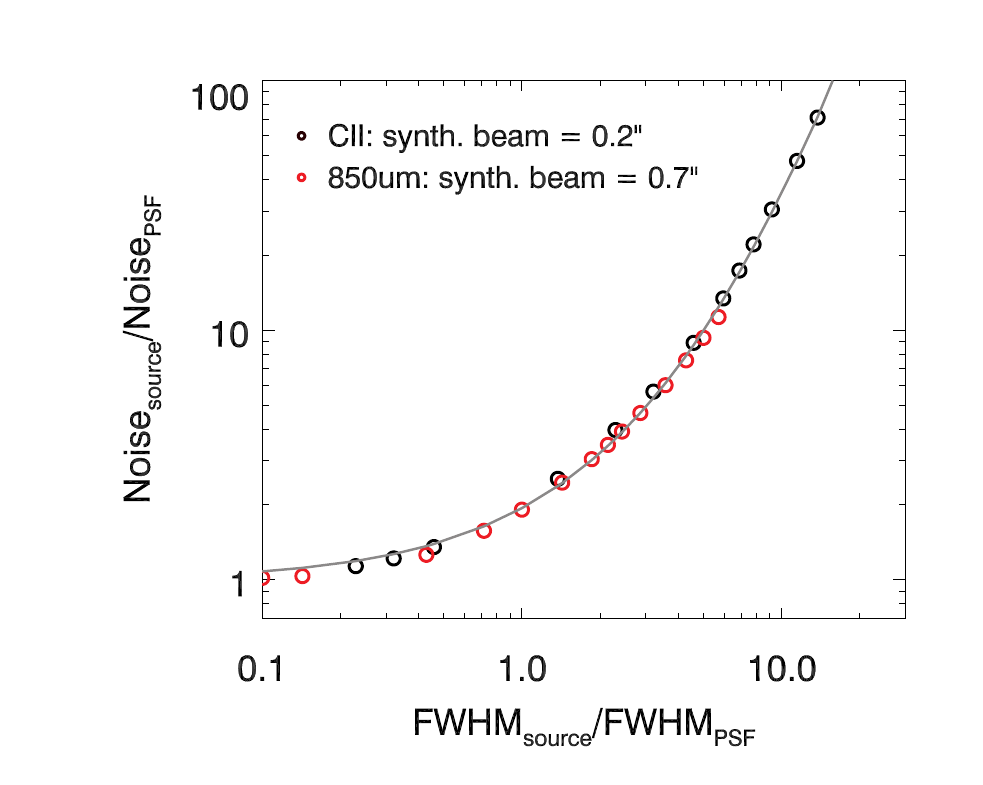}
\caption{Analysis of how the flux uncertainty (noise) changes when a source is resolved,
  with respect to the unresolved case. The noise obtained fitting an
  emission line with a Gaussian model normalized by that retrieved
  with a PSF fit is reported as a function of the source's size
  normalized by that of the beam. The fit of the datapoints  is
  reported gray solid line. Normalized in this way, the trend is independent on the resolution of the observations and 
  expected to hold quite generally for ALMA observations, at least at first order (to a second order, it should depend on the exact baseline distributions in the observing configuration).}
\label{fig:sims}
\end{figure}

From the \textit{HST} optical images we estimated that our sample galaxies
have FWHM sizes of $\sim$ 0.7 -- 1'' (see Section \ref{subsec:sample}). Since we wanted to measure total
\cii~ fluxes, we asked for ALMA observations choosing the configuration
C32-1, to get a resolution of $\sim$ 1''. However, the data were taken
with the configuration C32-3 instead, providing a $\sim$ 0.2''
resolution, higher than needed, and a maximum recoverable scale of
$\sim$ 3.5'': our galaxies are then spatially resolved. We tested
the impact of the resolution on our flux and size estimates as follows. With the
CASA task \texttt{simobserve} we simulated 2D Gaussians with
increasing FWHM (in the range 0.1'' -- 2''), mimicking observations
taken with $\sim$ 0.2'' resolution. We then fitted these mock data in
the \textit{uv} plane with the task \texttt{uvmodelfit}. Both sizes and fluxes are very well recovered even for large galaxies provided the data had large enough S/N.
Very similar results are obtained when simulating sources with GILDAS instead of CASA.

Although fluxes and sizes are well estimated when fitting the
emission lines in the \textit{uv} plane almost independently of the adopted
ALMA configuration, the S/N of the observations dramatically
decreases when the sources are resolved. To quantify it, we considered
the galaxy in our sample showing the highest S/N \cii~ emission line
(ID9347). We fitted its velocity-integrated map multiple times with the GILDAS algorithm, first with a point source model and then adopting a
Gaussian profile with increasing FWHM. In the following we call ``noise'' the uncertainty associated with the flux, as estimated by GILDAS during the fitting procedure. In Figure \ref{fig:sims} we
illustrate how the noise of an extended source changes
when it is resolved out. The noise is estimated as the uncertainty
associated with the flux, when fitting the data. We repeated the exercise for both the \cii~
emission line (synthesised beam $\sim$ 0.2'') and the 850 $\mu$m
continuum (synthesised beam $\sim$ 0.7''). By fitting the datapoints with a polynomial curve
we obtained the following relation:
\begin{equation}
y = 1.00 + 0.79x + 0.14x^2 + 0.01x^3
\label{eq:noise}
\end{equation}
where $y$ is the ratio of the source and PSF uncertainties ($y =
\mathrm{Noise_{source}}/\mathrm{Noise_{PSF}}$), and $x$ is the ratio of their
FWHM ($x = \mathrm{FWHM_{source}}/\mathrm{FWHM_{PSF}}$).

 Figure \ref{fig:sims} might be of
particular interest when proposing for observations, since the ALMA calculator
only provides sensitivity estimates assuming that the source is
unresolved. Our plots allow to rescale the sensitivity computed by the
calculator on the basis of the actual FWHM of the target, and therefore to
estimate the correct S/N to be expected in the observations. We notice however
that these predictions assume that the correct position and FWHM of
the source are known.

\section{B. Astrometry}
\label{app:astrometry}

When comparing our optical data with the observations of the \cii~
emission lines together with the 450 $\mu$m (Band 9) and 850 $\mu$m
(Band 7) continuum, there is an astrometric offset between \textit{HST} and
ALMA images. Considering only the galaxies with a line and/or
continuum detection (S/N $>$ 3), we estimated the average offsets
needed to align the luminosity peak of the \textit{HST} and ALMA datasets. We measured a systematic
shift of the \textit{HST} centroid with respect to the ALMA data of
$\sim -$ 0.2'' in declination and a non significant, negligible offset of $\sim$ 0.06'' in right ascension. 

We acknowledge that the astrometry offsets between \textit{HST} and
ALMA datasets in GOODS-S are a known issue (\citealt{Dunlop2017}, \citealt{Rujopakarn2016},
\citealt{Barro2016}, \citealt{Aravena2016b}, \citealt{Cibinel2017}). Our estimate is consistent with the ones reported in the
literature. A detailed map of the astrometry offset of the \textit{HST} imaging in GOODS-S will be provided by \cite{Dickinson2017}.

In our analysis we adopt the following coordinate shifts $\Delta RA = 0$'', $\Delta DEC = -0.2$''.

\section{C. Literature data}
\label{app:literature}

We briefly describe the literature samples that we used to complement
our observations and the methods used to derive the parameters considered in our analysis (redshift, \cii~
luminosity, IR luminosity, CO luminosity, molecular gas mass, specific
star formation rate, and metallicity). To properly compare different samples we converted all metallicity estimates to the calibration by \cite{Pettini2004} using the parametrizations by \cite{Kewley2008}. Also we homogenized all the IR luminosities reporting them to the 8 -- 1000 $\mu$m range.
\begin{itemize}[leftmargin=*]
\item \textit{Local dwarf galaxies} (\citealt{Cormier2015},
  \citealt{Cormier2017}). Sample of local dwarf galaxies observed
  with \textit{Herschel}/PACS and SPIRE as part of the DGS survey
  \citep{Madden2013}. They have metallicity ranging from $\sim$ 1/40
  Z$_\odot$ to near solar, SFR from $\sim 5\times 10^{-4}$ M$_\odot$
  yr$^{-1}$ to 25 M$_\odot$ yr$^{-1}$ and they are all nearby (maximum
  distance $\sim$ 200 Mpc). In this work we only consider the galaxies that have been followed-up with ATNF Mopra 22--m, APEX, and
  IRAM 30--m telescopes and show a CO(1-0)  emission line detection (\citealt{Cormier2014},
  \citealt{Cormier2017}, \citealt{deVis2017}). 
 We converted the CO luminosity of these
  sources (\citealt{Cormier2014}, \citealt{Cormier2017}) into molecular gas mass by
  using a conversion factor that depends on metallicity
  ($\alpha_{\mathrm{CO}} \sim Z^{-1.5}$, see Section
  \ref{subsec:metall}). Their IR luminosity was estimated fitting the
  IR SEDs with semi-empirical models \citep{Galliano2011}. \cite{Remy-Ruyer2014} estimated their SFR
  from the total infrared luminosity using the equation from
  \cite{Kennicutt1998} and their stellar mass from the 3.6 and 4.5
  $\mu$m flux densities using the formula of
  \cite{Eskew2012}. 

\item \textit{Local main-sequence galaxies} \citep{Accurso2017}. Sample of 
  intermediate mass ($9 < \log M_\star/M_\odot < 10$), local galaxies
  from the xCOLD GASS survey \citep{Saintonge2017} with metallicities
  in the range $0.4 < Z/Z_\odot < 1.0$. They have \textit{Herschel}
  \cii~ and IRAM CO(1-0) observations, together with auxiliary data
  from GALEX, WISE, and SDSS. \cite{Accurso2017} computed the molecular gas
  masses from the CO luminosity, considering a conversion factor that depends on metallicity ($\alpha_{\mathrm{CO}} \sim Z^{-1.5}$, see Section \ref{subsec:metall}). 
\cite{Saintonge2016} measured the SFR of these sources
  from  the combination of UV and IR photometry and their stellar mass
  from SDSS photometry. 

\item \textit{Local main-sequence galaxies} \citep{Stacey1991}. Sample of 
  local galaxies with KAO observations. We excluded those that were
  classified as starbursts (on the basis of their dust temperature:
  $T_\mathrm{dust} \geq 40$ K) and considered only the 6 ``normal''
  star-forming ones. CO observations taken with a similar beam size to the \cii~ ones are reported by \cite{Stacey1991}. We estimate the molecular gas mass for these galaxies considering a Milky-Way like $\alpha_{\mathrm{CO}} = 4.4$ K km s$^{-1}$ pc$^2$ conversion factor. Measurements of stellar masses, and metallicity are not
  available. In Figure \ref{fig:cii_ssfr} we report the average
  \cii -to-IR luminosity ratio of these 6 sources considering that they
  are in main-sequence ($sSFR/sSFR_{\mathrm{MS}} = 1$).

\item \textit{Local main-sequence and starburst galaxies} \citep{Brauher2008}. Sample of
  local galaxies observed with \textit{ISO}/LWS including
  ``normal'' star-forming systems, starbursts, and AGNs. In this
  analysis we only considered the 74 sources with both \cii~ and IR detection. The IR luminosity was estimated from the 25 $\mu$m, 60 $\mu$m, and 100 $\mu$m fluxes as reported by \cite{Brauher2008}.
 Molecular gas and stellar mass, and metallicity measurements are not available.

\item \textit{Local starbursts} (\citealt{Diaz-Santos2013},
  \citealt{Diaz-Santos2017}). Sample of local luminous infrared
  galaxies observed with \textit{Herschel}/PACS as part of GOALS
  \citep{Armus2009}. They have far-infrared luminosities in the range
  $2\times 10^9$ L$_\odot$ -- $2\times 10^{12}$ L$_\odot$ and sSFR $5\times10^{-12}$
  -- $3\times 10^{-9}$ yr$^{-1}$. 
No measurements of their molecular gas mass are available from the
literature. We therefore estimated $M_{\rm mol}$ considering the
models by \cite{Sargent2014} and \cite{Scoville2017} that parametrize
the dependence of galaxies' depletion time on their sSFR (see Section
\ref{subsec:cii_ssfr} for more details). The IR luminosity was estimated from the 60 $\mu$m and 100 $\mu$m as reported by \cite{Diaz-Santos2013}. Their SFR is estimated from IR luminosity \citep{Kennicutt1998} and their stellar
  mass from the IRAC 3.6 $\mu$m and Two Micron All Sky Survey (2MASS)
K-band photometry \citep{Howell2010}. The metallicity of these sources is not available. 

\item \textit{Redshift $z \sim 0.2$ Lyman break analogs}
  \citep{Contursi2017}. Sample of Lyman break analogs (namely,
  compact galaxies with UV luminosity $L_{\mathrm{UV}} > 2\times
  10^{10}$ L$_\odot$ and UV surface brightness $I_{\mathrm{1530\AA}} >
  10^9$ L$_\odot$ kpc$^{-2}$) at redshift 0.1 -- 0.3, with
  \textit{Herschel}/PACS \cii~ and IRAM CO(1-0) observations. Their IR luminosity was derived by fitting the IR SEDs of these sources with \cite{Draine2007} models. Their
  SFRs span the range 3 -- 100 M$_\odot$ yr$^{-1}$ and their sSFR are
  comparable to those of $z \sim 2$ main-sequence galaxies.  
We determined their molecular gas mass from the CO luminosity,
  using a conversion factor that depends on metallicity ($\alpha_{\mathrm{CO}} \sim Z^{-1.5}$, see Section \ref{subsec:metall}). 
  Their SFR has been derived from the IR luminosity considering the equation from \cite{Kennicutt1998}
  and their stellar masses from rest-frame optical photometry
  \citep{Overzier2009}.

\item \textit{Redshift $z \sim 0.5$ starbursts} \citep{Magdis2014}. Sample of
  (ultra)-luminous infrared galaxies at redshift 0.21 -- 0.88
  observed with \textit{Herschel}. They have an IR luminosity
  $L_{\mathrm{IR}} > 10^{11.5}$ L$_\odot$. Among them, 5 are
  classified as AGN host, QSO, or composite systems from optical or
  IRS data. 
 The gas mass
  has been estimated from the CO luminosity considering a conversion factor that depends on metallicity ($\alpha_{\mathrm{CO}} \sim Z^{-1.5}$, see Section \ref{subsec:metall}). Their IR luminosity was estimated fitting the IR SEDs of these sources with \cite{Draine2007} models. The SFR of these
  sources is derived from the IR luminosity considering the equation
  from \cite{Kennicutt1998}.

\item \textit{Redshift z = 1.8 lensed galaxy}
  \citep{Ferkinhoff2014}. Single galaxy lensed by the foreground galaxy observed with \textit{Herschel} and CSO/ZEUS. The gas mass of this galaxy was determined from the CO luminosity considering a conversion factor $\alpha_{\mathrm{CO}} = 4.4$ M$_\odot$ (K km s$^-1$ pc$^2$)$^{-1}$. Its IR luminosity was estimated fitting the IR SED with \cite{Siebenmorgen2007} models. In the following we report the unlensed luminosities.

\item \textit{Redshift z = 1.8 main-sequence galaxies}
  \citep{Brisbin2015}. Sample of galaxies at redshift $z \sim 1.8$
  observed with CSO/ZEUS. The observed IR luminosity of these
  sources ranges between $7\times 10^{11}$ L$_\odot$ -- $6\times
  10^{12}$ L$_\odot$ and was estimated fitting the IR SED with the models by \cite{Dale2002}. Measurements of 
  molecular gas masses are not available. The star formation rate has
  been estimated from the IR luminsity, considering the equation from
  \cite{Kennicutt1998}. The stellar mass has been estimated from the 2
  $\mu$m IRAC flux \citep{Magdis2010}. Metallicity measurements are
  not available. We note that some of the galaxies by \citet{Brisbin2015} might be
  lensed. While the L$_{\mathrm{\cii}}$/L$_{\mathrm{IR}}$ ratio should
  not be particularly affected by differential magnification, the
  absolute \cii~ and IR luminosities might instead be amplified.

\item \textit{Redshift z = 2 lensed main-sequence galaxy}
  \citep{Schaerer2015}. Single galaxy lensed by the foreground galaxy cluster MACS J0451+0006 observed with \textit{HST}, \textit{Spitzer}, \textit{Herschel}, PdBI, and ALMA. The gas mass of this galaxy was determined from the CO luminosity considering a conversion factor that depends on metallicity ($\alpha_{\mathrm{CO}} \sim Z^{-1.5}$, see Section \ref{subsec:metall}). \cite{Dessauges-Zavadsky2015} estimated its IR luminosity fitting the IR SED with \cite{Draine2007} models and derived its SFR and stellar mass from the best energy conserving SED fits, obtained under the hypothesis of an extinction fixed at the observed IR-to-UV luminosity ratio following the prescriptions of \cite{Schaerer2013}.

\item \textit{Redshift $z \sim 1$ -- 2 main-sequence and starbursts}
  \citep{Stacey2010}. Sample of galaxies at redshift 1 -- 2
  observed with CSO/ZEUS. The observed far-IR luminosity of these
  sources ranges between $3\times 10^{12}$ L$_\odot$ -- $2.5\times
  10^{14}$ L$_\odot$, although two of them are lensed. In the
  following we report the observed luminosities since the magnification
  factors are generally very uncertain or unknown. Both AGN and
  star-forming galaxies are included. Measurements of 
  molecular gas masses are not available, as well as estimates of the
  sources' stellar mass, and metallicity. The IR luminosity was estimated from the 12 $\mu$m, 25 $\mu$m, 60 $\mu$m, and 100 $\mu$m fluxes as reported by \cite{Stacey2010}.

\item \textit{Redshift $z = 4.44$ main-sequence galaxy} \citep{Huynh2014}. Single galaxy observed with ATCA, ALMA, \textit{Herschel}, and \textit{HST}. The gas mass of this galaxy was determined from the CO luminosity considering a conversion factor that depends on metallicity ($\alpha_{\mathrm{CO}} \sim Z^{-1.5}$, see Section \ref{subsec:metall}). \cite{Huynh2014} estimated its IR luminosity fitting the IR SED with \cite{Chary2001} models. Its SFR was derived from the IR luminosity following the calibration from \cite{Kennicutt1998} and its stellar mass from the \textit{H}-band magnitude together with an average mass-to-light ratio for a likely sub-millimeter galaxy star formation history \citep{Swinbank2012}.

\item \textit{Redshift $z \sim 5.5$ main-sequence galaxies}
  \citep{Capak2015}. Sample of star-forming galaxies at redshift 5
  -- 6 observed with ALMA and \textit{Spitzer}. In the following we only report the 4 galaxies with detected \cii~ emission together
  with the average \cii~ luminosity obtained by stacking the 6
  non-detections. Two galaxies were also serendipitously detected in
  \cii~ and added to the sample. The SFRs range between 3 -- 169
  M$_\odot$ yr$^{-1}$ and the stellar masses $9.7 < \log
  M_\star/M_\odot < 10.8$. 
 CO observations are not available, so we estimated the molecular gas masses using the integrated Schmidt-Kennicutt relation for main-sequence galaxies reported by \cite{Sargent2014}. The IR luminosity was estimated using the
grey body models from \cite{Casey2012}. \cite{Capak2015} estimated
  the SFR of the sources by summing the UV and IR luminosity and the
  stellar mass by fitting SED models to the UV to IR photometry. The
  metallicity of these galaxies is not available.

\item \textit{Redshift $z \sim 2$ -- 6 lensed galaxies}
  \citep{Gullberg2015}. Sample of strongly lensed dusty
  star-forming galaxies in the redshift range 2.1 -- 5.7 selected from
  the South Pole Telescope survey (\citealt{Vieira2010},
  \citealt{Carlstrom2011}) on the basis of their 1.4 mm flux
  ($S_{\mathrm{1.4mm}} > 20$ mJy) and followed-up with ALMA and
  \textit{Herschel}/SPIRE. Among them, 11 sources also have low-$J$ CO
  detections from ATCA. In the following we report the de-magnified
  luminosities, where the magnification factors are taken from
  \citep{Spilker2016}. 
 The molecular gas mass has been computed
  considering the CO luminosity and an $\alpha_{\mathrm{CO}}$
  conversion factor derived for each source on the basis of their dynamical mass (see
\citealt{Aravena2016} for more details). The adopted $\alpha_{\mathrm{CO}}$ factors have values in the range 0.7 -- 12.3 M$_\odot$ K km s$^{-1}$ pc$^{-2}$. Their IR luminosity was estimated fitting the IR SEDs of these sources with greybody models from \cite{Greve2012}. The stellar mass, and metallicity of
  these galaxies are not available.
\end{itemize}

\onecolumn
\begin{longtable}{c c c c c c c c}
\caption{Compilation of literature data used in this paper. The full
  table is available online.}
\label{tab:acii}
%\begin{tabu}
\\
\toprule
\midrule
ID & z & L$_{\mathrm{\cii}}$ & L$_{\mathrm{IR}}$ & L'$_{\mathrm{CO}}$ & M$_{\mathrm{mol}}$ & sSFR & 12 + log(O/H) \\[3pt]
    &   &   [L$_\odot$]         &      [L$_\odot$]   &    [L$_\odot$]        &     [M$_\odot$]   & [yr$^{-1}$] &               \\[3pt]
\midrule
\multicolumn{8}{c}{Local dwarf galaxies (\citealt{Cormier2015}, \citealt{Cormier2017})} \\
\midrule
Haro11           & 0.021 & $1.3\times 10^8$ & $1.9\times 10^{11}$ & $9.8\times 10^7$ & $1.7\times 10^9$ & $1.4\times 10^{-9}$ & 8.30 \\
Haro2             & 0.005 & $1.4\times 10^7$ & $6.0\times 10^9$    & $4.1\times 10^7$ & $4.8\times 10^8$ & $2.2\times 10^{-10}$& 8.42 \\
Haro3             & 0.005 & $1.3\times 10^7$ & $5.2\times 10^9$    & $1.9\times 10^7$ & $4.3\times 10^8$ & $2.0\times 10^{-10}$& 8.22 \\
He2-10           & 0.002 & $1.1\times 10^7$ & $5.2\times 10^9$    & $3.0\times 10^7$ & $2.2\times 10^8$ & $1.7\times 10^{-10}$& 8.55 \\
IIZw40            & 0.003 & $1.9\times 10^6$ & $2.7\times 10^9$    & $1.6\times 10^6$ & $1.1\times 10^8$ & $2.8\times 10^{-9}$  & 7.92 \\
\bottomrule
%\end{tabu}
\end{longtable}
\begin{center}
\smallskip
\begin{minipage}{18cm}
\footnotesize
\textbf{Columns} (1) Galaxy ID; (2) Redshift; (3) \cii~ luminosity; (4) Infrared luminosity; (5) CO luminosity; (6) Molecular gas mass; (7) Specific star formation rate; (8) Gas-phase metallicity. \\
\textbf{Notes.} For the sample by \citet{Diaz-Santos2013}, \citet{Diaz-Santos2017} we report two molecular gas mass estimates. They have both been obtained considering the sSFR of each galaxy, their SFR, and the relation between depletion time and sSFR (\citealt{Sargent2014}, \citealt{Scoville2017}). The difference between the two estimates consists in the model that we assumed: the first is derived using the mean depletion time obtained averaging the parametrization by \citet{Sargent2014} and \citet{Scoville2017}, whereas the second is estimated considering only the model by \citet[see Section \ref{subsec:cii_ssfr} for a more detailed discussion]{Scoville2017}.
\end{minipage}
\end{center}

\end{document}